\documentclass[aps,twocolumn,superscriptaddress,a4paper,floatfix]{revtex4-2}
\usepackage[latin1]{inputenc}
\usepackage{graphicx}
\usepackage{amsmath,amssymb}
\usepackage{hyperref}

\newcommand \beq{\begin{eqnarray}}
\newcommand \eeq{\end{eqnarray}}

\begin{document}

\title{Comparative study for two-terminal transport through a lossy one-dimensional quantum wire}

\author{Shun Uchino}
\affiliation{Advanced Science Research Center, Japan Atomic Energy Agency, Tokai 319-1195, Japan}


\begin{abstract}
Motivated by realization of the dissipative quantum point contact in ultracold atomic gases,
we investigate a two-terminal mesoscopic transport system in which 
a single-particle loss is locally present in a one-dimensional chain.
By means of  the Dyson equation approach in the Keldysh formalism that can incorporate dissipative effects, 
we reveal analytic structures of the particle and energy currents whose formal expressions correspond to ones in certain three-terminal systems where the particle loss is absent.
The obtained formulas are also consistent with non-hermitian and three-terminal Landauer-B\"{u}ttiker analyses. 
The universality on the current expressions holds regardless of quantum statistics, and may be useful for understanding 
lossy two-terminal transport in terms of three-terminal transport and vice versa.

\end{abstract}


\maketitle

\section{Introduction}
Along with the intensive momentum of quantum technology, research on quantum simulation that extracts essential features of complex
quantum many-body problems develops at a rapid pace~\cite{schafer2020}. 
Since Feynman's proposal~\cite{feynman}, several promising quantum simulators such as ultracold atomic gases~\cite{bloch2012}, 
trapped ions~\cite{blatt2012}, photonic systems~\cite{aspuru2012}, and superconducting qubits~\cite{houck2012}
have been introduced and investigated a broad spectrum of quantum many-body problems ranging from condensed matter physics to
astrophysics, and high energy physics such as particle and nuclear physics~\cite{RevModPhys.86.153}.

Among them, ultracold atomic gases have reached a high level of maturity owing to controllability and flexibility of experiments.
As is often said, one of the advantages in ultracold atomic gases is that the systems can normally be treated as closed ones in that effects of
an uncontrollable external environment are negligible.
In fact, thermalization and non-thermalization problems occurring in closed quantum many-body systems have successfully been verified with the atomic gases~\cite{ueda2020}.
More recently,  the controllable atomic gases enable to explore
open quantum many-body systems as  well.
By applying various schemes to perturb the systems~\cite{syassen2008,PhysRevLett.110.035302,PhysRevLett.115.140402,tomita2017,PhysRevLett.121.200401,PhysRevA.100.053605,konishi2021}, 
one can engineer one-body, two-body, and three-body losses and analyze dissipation effects in quantum many-body systems.

How dissipation affects
quantum transport is an important theme in quantum technology~\cite{datta1997,nazarov}  and is now diagnosed 
in ultracold atomic gases.
The epochal realization is lossy Josephson junction arrays of Bose-Einstein condensates,
where the negative differential conductance~\cite{PhysRevLett.115.050601} and  bistability between superfluid and resistive 
states~\cite{PhysRevLett.116.235302,benary2022} have been found.
\begin{figure}[htbp]
\centering
\includegraphics[width=7cm]{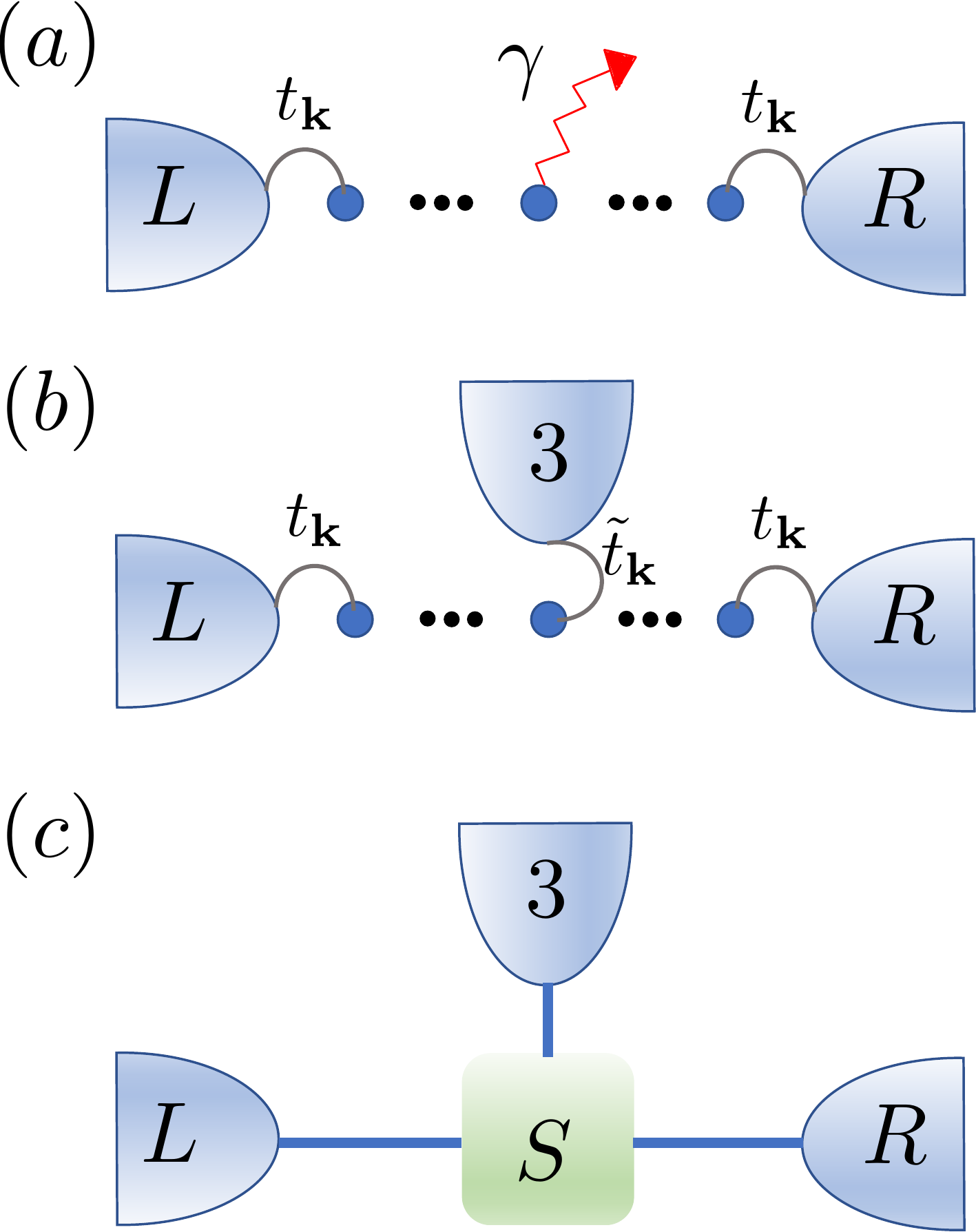}
\caption{\label{fig1}Mesoscopic transport systems discussed in this work.
(a) The lossy two-terminal transport system where
the single-particle loss occurs at the center site in the 1D chain.
(b) Three terminal transport system where the center site in the 1D chain couples to the third reservoir.
(c) Three terminal scattering problem where the sample region in the Landauer-B\"{u}ttiker formalism is characterized by an $S$-matrix. }
\end{figure}
Another striking realization, which is the subject of this work, is two-terminal transport via
the dissipative quantum point contact~\cite{PhysRevA.100.053605}, where a local single-particle loss is introduced inside the short one-dimensional wire
and leads to a conductance plateau less than $1/h$.
Although the dissipation effects discussed in ultracold atomic gases can usually be explained
in terms of the Lindblad master equation~\cite{daley2014}, unlike Bose-Einstein condensates where the quantum jump term being difficult to handle is greatly simplified
in the Gross-Pitaevskii analysis, a proper prescription in the quantum point contact system has yet to be fully understood.
In order to explain the experimental data of the dissipative quantum point contact, Ref.~\cite{PhysRevA.100.053605} 
introduced the Landauer-B\"{u}ttiker formalism with a complex potential in the belief that
the quantum jump term does not play an essential role in transport.
Whilst an analysis with a complex potential has indeed been utilized in the context of inelastic scattering problems~\cite{moiseyev2011,ashida2020},
 it may also be important to understand dissipative quantum point contact transport on the basis of the Lindblad formalism including the quantum jump term.

In this paper, 
we investigate two-terminal current flows through the lossy one-dimensional chain that is a model of the dissipative quantum point contact. 
We note that the similar model has recently been addressed in Refs.~\cite{visuri2022,PhysRevB.102.205131}.
Compared to these previous works, 
here we focus on analytic structures of the energy current as well as the particle current, and a connection to other approaches including the non-hermitian Landauer-B\"{u}ttiker formalism.
To this end,  we adopt an analysis based on Dyson equations in the Keldysh functional integral approach extended to open quantum systems~\cite{sieberer2016}, 
and reveal a relationship between such a dissipative system and
a similar system in which dissipation is absent but an additional reservoir is attached. 
By taking an appropriate limit shown below, we show that regardless of fermions or bosons
particle and energy currents in these two systems coincide with each other. 
In addition, the formal expressions of the currents are found to be consistent with ones based on
in the non-hermitian analysis obtained in Ref.~\cite{PhysRevA.100.053605} and in certain three-terminal Landauer-B\"{u}ttiker 
analysis.
What is remarkable is that formal current expressions of the lossy two-terminal systems 
can be  reproduced with three-terminal systems consisting of non-empty reservoirs, which paves the way to 
simulation of lossy two-terminal transport realized in ultracold atomic gases in terms of
lossless three-terminal transport realized in condensed matter systems and vice versa.

The paper is organized as follows.
Section II performs formulation for the systems of Fig.~1~(a) and~(b).
Section III and IV apply our formalism to single-site and multi-site cases, respectively.
Section V performs the three-terminal Landauer-B\"{u}ttiker analysis for the system of Fig.~1~(c).
Section VI gives summary and outlook on this work.
In Appendix, we give a derivation of the functional integral form of the partition function whose density matrix
obeys the Lindblad master equation.

\section{Formulation of the problem}
We consider a two-terminal mesoscopic transport system of noninteracting particles that can be either bosons or fermions. 
In this work, we focus on a situation that a one-dimensional tight-binding chain is attached between two macroscopic reservoirs.
By adopting units of $\hbar=k_B=1$, the corresponding Hamiltonian is given by
\beq
&&\hat{H}=\hat{H}_L+\hat{H}_R+\hat{H}_{1D}+\hat{H}_T, \label{eq:dissipative-h}\\
&&\hat{H}_{1D}=\sum_{i=-N}^N\epsilon_i \hat{d}_i^{\dagger}\hat{d}_i -\sum_{i=-N}^{N-1}t_i(\hat{d}^{\dagger}_i\hat{d}_{i+1}+\hat{d}^{\dagger}_{i+1}\hat{d}_i) \\
&&\hat{H}_T=-\sum_{\mathbf{k}}\Big[t_{\mathbf{k}}\hat{\psi}^{\dagger}_{L,\mathbf{k}}\hat{d}_{-N}+t_{\mathbf{k}}\hat{\psi}^{\dagger}_{R,\mathbf{k}}\hat{d}_N\Big]+\text{h.c.} 
\eeq
Here, $\hat{H}_{L(R)}=\sum_{\mathbf{k}}(\epsilon_{\mathbf{k}}-\mu_{L(R)} )\hat{\psi}^{\dagger}_{L(R),\mathbf{k}}\hat{\psi}_{L(R),\mathbf{k}}$,
where $\epsilon_{\mathbf{k}}=\frac{\mathbf{k}^2}{2m}$ with  particle mass $m$,  
is the Hamiltonian of the left(right) reservoir measured from the chemical potential $\mu_{L(R)}$,
$\hat{H}_{1D}$ is the Hamiltonian of the one-dimensional chain with nearest neighbor hopping $t_i$, and $\hat{H}_T$ is the coupling between the reservoirs and the one-dimensional chain that is modeled as the tunneling term
with tunneling amplitude $t_{\mathbf{k}}$.
For a symmetric reason,  we assume that the onsite potential energy $\epsilon_i$ has the property $\epsilon_i=\epsilon_{-i}$,
the hopping amplitude satisfies $t_{-i}=t_{i-1}$ for $i=0,1,\cdots, N$,
and the site number of the 1D chain $L$ is set to be $L=2N+1$.
In addition, $\hat{\psi}_{L(R)}$ and $\hat{d}$ are respectively annihilation field operators  in reservoir $L(R)$ and one-dimensional chain.

For a comparative purpose, we also introduce a three-terminal system described by the following Hamiltonian:
\beq
&&\hat{\tilde{H}}=\hat{H}_L+\hat{H}_R+\hat{H}_{1D}+\hat{\tilde{H}}_T+\hat{H}_3,\label{eq:three-h} \\
&&\hat{\tilde{H}}_T=\hat{H}_T-\sum_{\mathbf{k}}\tilde{t}_{\mathbf{k}}\hat{\psi}^{\dagger}_{3,\mathbf{k}}\hat{d}_{0}+\text{h.c.} 
\eeq
Here $\hat{H}_3$ is an additional reservoir Hamiltonian that is similar to $\hat{H}_{L}$ and $\hat{H}_R$
and is expressed with the field operator $\hat{\psi}_{3,\mathbf{k}}$.
In addition, we assume that a coupling to the third reservoir with tunnel amplitude $\tilde{t}_{\mathbf{k}}$ occurs at the center site in the 1D chain. 

We now come back to Eq.~\eqref{eq:dissipative-h} and  introduce dissipation such that  a single-particle loss is present at the center site in the chain.
Inspired by the dissipative quantum point contact experiment~\cite{PhysRevA.100.053605}, we assume that the system obeys
 the following Lindblad Master equation~\cite{breuer2002}:
\beq
\partial_{\tau}\hat{\rho}=-i[\hat{H},\hat{\rho}]+\gamma\Big(\hat{d}_0\hat{\rho} \hat{d}^{\dagger}_0-\frac{\{\hat{d}^{\dagger}_0\hat{d}_0,\hat{\rho}\}}{2}\Big),
\label{eq:lindblad-loss}
\eeq
where $\hat{\rho}$ is the density matrix operator and $\gamma$ is the dissipation rate.
For $\gamma=0$, this equation is reduced to the  von Neumann equation describing unitary time evolutions.
In the presence of nonzero $\gamma$, additional terms describing dissipation effects come out.
It is also convenient to introduce the adjoint equation.
For a Heisenberg operator $\hat{A}$, the corresponding adjoint equation is written as~\cite{breuer2002}
\beq
\partial_{\tau}\hat{A}=i[\hat{H},\hat{A}]+\gamma\Big[\hat{d}^{\dagger}_0\hat{A}\hat{d}_0-\frac{\{\hat{d}^{\dagger}_0\hat{d}_0,\hat{A}\}}{2} \Big].
\eeq
By using above, the operator of particle number growth rate in each reservoir is given by
\beq
\partial_{\tau}\hat{N}_{L(R)}=i\sum_{\mathbf{k}}t_{\mathbf{k}}\hat{\psi}^{\dagger}_{L(R),\mathbf{k}}\hat{d}_{-N(N)}+\text{h.c.},
\eeq
where we use
\beq
[\hat{N}_{L(R)},\hat{d}_0]=[\hat{N}_{L(R)},\hat{d}^{\dagger}_0]=0.
\eeq
Similarly, the operator of energy absorption rate in each reservoir is given by
\beq
\partial_{\tau}\hat{H}_{L(R)}+\mu_{L(R)}\partial_{\tau}\hat{N}_{L(R)}=i\sum_{\mathbf{k}}t_{\mathbf{k}}\epsilon_{\mathbf{k}}\hat{\psi}^{\dagger}_{L(R)}\hat{d}_{-N(N)}+\text{h.c.},
\nonumber\\
\eeq
where we note that $\mu_{L(R)}\partial_{\tau}\hat{N}_{L(R)}$ is added in order to describe the energy flow.
By using these operators, the particle and energy current operators flowing between reservoirs are respectively expressed as
\beq
&&\hat{I}=\frac{\partial_{\tau}(-\hat{N}_L+\hat{N}_R)}{2},\\
&&\hat{I}_E=\frac{\partial_{\tau}(-\hat{H}_L-\mu_L\hat{N}_L+\hat{H}_R+\mu_R\hat{N}_R)}{2},
\eeq
where of course $\mu_{L(R)}$ is assumed to be a constant in time~\footnote{In a precise sense, ultracold atomic gases are finite systems and therefore
the chemical potential in each reservoir is not exactly a constant in time.
Thus, the constant $\mu_{L(R)}$ treatment becomes a good approximation if
 thermalization in each reservoir is enough fast compared to the transport (relaxation) time scale~\cite{krinner2017}.}.

In order to evaluate averages of operators defined above, we harness
 functional integral formulation of the Keldysh formalism~\cite{kamenev2011}.
There, we can consider the following partition function~\cite{sieberer2016} (see also Appendix):
\beq
Z=\int {\cal D}[\bar{\psi},\psi,\bar{d},d]  e^{iS_L+iS_R+iS_{1D}+iS_T+iS_{\text{loss}}}, \label{eq:partition} 
\eeq
where
\begin{widetext}
\beq
S_{L(R)}&&=\int_{-\infty}^{\infty}d\tau\Big[\bar{\psi}^+_{L(R)}i\partial_{\tau}\psi^+_{L(R)}-\bar{\psi}^-_{L(R)}i\partial_{\tau}\psi^-_{L(R)} -H^+_{L(R)}+H^-_{L(R)} \Big],
\eeq
\beq
S_{1D}&&=\int_{-\infty}^{\infty}d\tau\Big[\sum_{i=-N}^N\{\bar{d}^+_i(i\partial_{\tau} -\epsilon_i )d^+_i-\bar{d}^-_i(i\partial_{\tau}-\epsilon_i)d^-_i\} +\sum_{i=-N}^{N-1}t_i\{\bar{d}^+_id^+_{i+1}+
\bar{d}^+_{i+1}d^+_{i}-\bar{d}^-_id^-_{i+1}-\bar{d}^-_{i+1}d^-_{i}\}  \Big],
\eeq
\beq
S_T&&=\int_{-\infty}^{\infty}d\tau\sum_{\mathbf{k}}\Big[t_{\mathbf{k}}\bar{\psi}^+_{L,\mathbf{k}}d^+_{-N}+t^*_{\mathbf{k}}\bar{d}^+_{-N}\psi^+_{L,\mathbf{k}} 
+t_{\mathbf{k}}\bar{\psi}^+_{R,\mathbf{k}}d^+_N+t^*_{\mathbf{k}}\bar{d}^+_N\psi^+_{R,\mathbf{k}}   \nonumber\\
&& -t_{\mathbf{k}}\bar{\psi}^-_{L,\mathbf{k}}d^-_{-N}- t^{*}_{\mathbf{k}}\bar{d}^-_{-N}\psi^-_{L,\mathbf{k}} 
 -t_{\mathbf{k}}\bar{\psi}^-_{R,\mathbf{k}}d^-_N- t^*_{\mathbf{k}}\bar{d}^-_N\psi^-_{R,\mathbf{k}}  \Big],
 \eeq
 \beq
S_{\text{loss}}&&=\int_{-\infty}^{\infty}d\tau i\gamma\Big[\frac{\bar{d}^+_0d^+_0+\bar{d}^-_0d^-_0}{2}-\bar{d}^-_0d^+_0\Big].
\eeq
Here, superscripts + and - represent forward and backward contours, respectively,
and $H^{\pm}_{L(R)}=\sum_{\mathbf{k}}(\epsilon_{\mathbf{k}}-\mu_{L(R)})\bar{\psi}^{\pm}_{L(R),\mathbf{k}}\psi^{\pm}_{L(R),\mathbf{k}}$.
Although the expressions above are available regardless of bosons or fermions, 
we note that the fields are expressed with complex numbers for bosons and Grassmann numbers for fermions.
As can be seen from the expressions above, an advantage of this formalism is that 
the seemingly complicated dissipator terms that consist of non-hermitian and quantum jump terms can be replaced by
the quadratic action term $S_{\text{loss}}$.

In place of the representation based on the forward and backward counters, 
we convert to the following rotated representation~\cite{kamenev2011}:
\beq
\begin{pmatrix}
\psi^{cl}\\
\psi^q
\end{pmatrix}
=\frac{1}{\sqrt{2}}
\begin{pmatrix}
 1 & 1 \\
1 & -1
\end{pmatrix}
\begin{pmatrix}
\psi^{+}\\
\psi^-
\end{pmatrix}, \ \ 
\begin{pmatrix}
d^{cl}\\
d^q
\end{pmatrix}
=\frac{1}{\sqrt{2}}
\begin{pmatrix}
 1 & 1 \\
1 & -1
\end{pmatrix}
\begin{pmatrix}
d^{+}\\
d^-
\end{pmatrix}, 
\label{eq:transformation}
\eeq
\beq
\begin{pmatrix}
\bar{\psi}^{cl}\\
\bar{\psi}^q
\end{pmatrix}
=\frac{1}{\sqrt{2}}
\begin{pmatrix}
 1 & 1 \\
1 & -1
\end{pmatrix}
\begin{pmatrix}
\bar{\psi}^{+}\\
\bar{\psi}^-
\end{pmatrix}, \ \ 
\begin{pmatrix}
\bar{d}^{cl}\\
\bar{d}^q
\end{pmatrix}
=\frac{1}{\sqrt{2}}
\begin{pmatrix}
 1 & 1 \\
1 & -1
\end{pmatrix}
\begin{pmatrix}
\bar{d}^{+}\\
\bar{d}^-
\end{pmatrix}.
\label{eq:bar}
\eeq
Under this transformation, 
 the reservoir action is expressed as
\beq
S_{L(R)}
=\sum_{\mathbf{k}}\int_{-\infty}^{\infty} d\tau \int_{-\infty}^{\infty}d\tau'
\begin{pmatrix}
 \bar{\psi}^{cl}_{L(R)}  &
\bar{\psi}^{q}_{L(R)}
\end{pmatrix}_{\tau}
\begin{pmatrix}
0 & [g^{-1}_{L(R)}(\mathbf{k},\tau-\tau')]^{A} \\
[g^{-1}_{L(R)}(\mathbf{k},\tau-\tau') ]^{R}  &  [g^{-1}_{L(R)}(\mathbf{k},\tau-\tau') ]^{K}
\end{pmatrix}
\begin{pmatrix}
\psi^{cl}_{L(R)}\\
\psi^q_{L(R)}
\end{pmatrix}_{\tau'},
\eeq
where 
$[g^{-1}]^{R(A)}$ and $[g^{-1}]^{K}$ are the retarded (advanced) and Keldysh components in inverse Green's function, respectively.
In addition, $S_{\text{loss}}$ is transformed into
\beq
S_{\text{loss}}=\int_{-\infty}^{\infty} d\tau
\begin{pmatrix}
 \bar{d}^{cl}_{0}  &
\bar{d}^{q}_{0}
\end{pmatrix}
\begin{pmatrix}
0 & -\frac{i\gamma}{2} \\
\frac{i\gamma}{2}  &  i\gamma
\end{pmatrix}
\begin{pmatrix}
d^{cl}_{0}\\
d^q_{0}
\end{pmatrix}.
\eeq
Thus, non-infinitesimal imaginary parts appear in inverse Green's function at the center site.
It is also instructive to note that the conventional representation for fermions is 
the so-called Larkin-Ovchinnikov one where $\psi$ and $d$ fields transform in the same way as Eq.~\eqref{eq:transformation}, while
$\bar{\psi}$ and $\bar{d}$ fields transform in the following way~\cite{kamenev2011}: 
$
\begin{pmatrix}
\bar{\psi}^{1}\\
\bar{\psi}^2
\end{pmatrix}
=\frac{1}{\sqrt{2}}
\begin{pmatrix}
 1 & -1 \\
1 & 1
\end{pmatrix}
\begin{pmatrix}
\bar{\psi}^{+}\\
\bar{\psi}^-
\end{pmatrix}, $ and
$
\begin{pmatrix}
\bar{d}^{1}\\
\bar{d}^2
\end{pmatrix}
=\frac{1}{\sqrt{2}}
\begin{pmatrix}
 1 & -1 \\
1 & 1
\end{pmatrix}
\begin{pmatrix}
\bar{d}^{+}\\
\bar{d}^-
\end{pmatrix}.
$
In this paper, however, we stick to use Eq~\eqref{eq:bar}, 
which allows us to discuss both bosonic and fermionic transport in a unified manner.

We are interested in steady transport in which average currents do not depend on time.
In this case, it is convenient to move onto  the frequency space.
By taking into account the fact that origins of frequencies between reservoirs are shifted by a chemical potential difference $\Delta\mu\equiv \mu_L-\mu_R$,
the reservoir action in the frequency space is given by
\beq
S_{L/R}
=\sum_{\mathbf{k}}\int_{-\infty}^{\infty} \frac{d\omega}{2\pi}
\begin{pmatrix}
 \bar{\psi}^{cl}_{L/R}  &
\bar{\psi}^{q}_{L/R}
\end{pmatrix}
\begin{pmatrix}
0 & [g^{-1}_{L/R}(\mathbf{k},\omega\mp\Delta\mu/2)]^{A} \\
[g^{-1}_{L/R}(\mathbf{k},\omega\mp\Delta\mu/2) ]^{R}  &  [g^{-1}_{L/R}(\mathbf{k},\omega\mp\Delta\mu/2) ]^{K}
\end{pmatrix}
\begin{pmatrix}
\psi^{cl}_{L/R}\\
\psi^q_{L/R}
\end{pmatrix}.
\label{eq:uncoupled}
\eeq
Here, 
the retarded and advanced components are determined as
\beq
[g^{-1}(\mathbf{k},\omega)]^{R/A}=\omega-\xi_{\mathbf{k}}\pm i0^+.
\eeq
Since we deal with the two-terminal transport system in which two macroscopic reservoirs are in thermal equilibrium 
where the fluctuation-dissipation relation holds in each reservoir~\cite{kamenev2011},  $[g^{-1}]^{K}$ obeys
\beq
[g^{-1}(\mathbf{k},\omega)]^{K}=\Big([g^{-1}(\mathbf{k},\omega)]^{R}-[g^{-1}(\mathbf{k},\omega)]^{A} \Big)[1\pm 2n(\omega)],\nonumber\\
\eeq
where $n(\omega)=\frac{1}{e^{\omega/T}\mp 1}$ with temperature $T$ is the distribution function and
upper and lower signs are for bosons and fermions, respectively.

On the other hand, the partition function in the three terminal system described by Eq.~\eqref{eq:three-h}
is straightforwardly obtained as
\beq
\tilde{Z}=\int {\cal D}[\bar{\psi},\psi,\bar{d},d]  e^{iS_L+iS_R+iS_{1D}+i\tilde{S}_T+S_3}, 
\eeq
where
\beq
S_{3}&&=\int_{-\infty}^{\infty}d\tau\Big[\bar{\psi}^+_{3}i\partial_{\tau}\psi^+_{3}-\bar{\psi}^-_{3}i\partial_{\tau}\psi^-_{3} -H^+_{3}+H^-_{3} \Big],\\
\tilde{S}_T&&=S_T+\int_{-\infty}^{\infty}d\tau\sum_{\mathbf{k}}\Big[\tilde{t}_{\mathbf{k}}\bar{\psi}^+_{3,\mathbf{k}}d^+_0+\tilde{t}^*_{\mathbf{k}}\bar{d}^+_0\psi^+_{3,\mathbf{k}} 
 -\tilde{t}_{\mathbf{k}}\bar{\psi}^-_{3,\mathbf{k}}d^-_0- \tilde{t}^{*}_{\mathbf{k}}\bar{d}^-_0\psi^-_{3,\mathbf{k}}   \Big].
\eeq

\end{widetext}

For calculation of physical quantities, one need to evaluate Green's functions. 
In fact, it is easy to confirm that the  particle and energy currents are respectively expressed as
\beq
&&I(\tau)=\frac{\sum_{\mathbf{k}}\text{Re}\Big[\pm t_{\mathbf{k}}G^K_{dL}(\mathbf{k},\tau,\tau) \mp t_{\mathbf{k}}G^K_{dR}(\mathbf{k},\tau,\tau)\Big]}{2}
\nonumber\\
&& =\int_{-\infty}^{\infty}\frac{d\omega}{4\pi}\sum_{\mathbf{k}}\text{Re}\Big[\pm t_{\mathbf{k}}G^K_{dL}(\mathbf{k},\omega) \mp t_{\mathbf{k}}G^K_{dR}(\mathbf{k},\omega)\Big],
\eeq
\beq
&&I_E(\tau)=\frac{\sum_{\mathbf{k}}\epsilon_{\mathbf{k}}\text{Re}\Big[\pm t_{\mathbf{k}}G^K_{dL}(\mathbf{k},\tau,\tau) \mp
t_{\mathbf{k}}G^K_{dR}(\mathbf{k},\tau,\tau)\Big]}{2}\nonumber\\
&&=\int_{-\infty}^{\infty}\frac{d\omega}{4\pi}\sum_{\mathbf{k}}\epsilon_{\mathbf{k}}\text{Re}\Big[\pm t_{\mathbf{k}}G^K_{dL}(\mathbf{k},\omega) \mp t_{\mathbf{k}}G^K_{dR}(\mathbf{k},\omega)\Big].
\eeq
Here
\beq
G^{K}_{dL}(\mathbf{k},\tau,\tau')=-i\langle [\hat{d}_{-N}(\tau),\hat{\psi}^{\dagger}_{L,\mathbf{k}}(\tau')]_{\pm} \rangle,\\
G^{K}_{dR}(\mathbf{k},\tau,\tau')=-i\langle [\hat{d}_{N}(\tau),\hat{\psi}^{\dagger}_{R,\mathbf{k}}(\tau')]_{\pm} \rangle,
\eeq
and upper and lower signs are again for bosons and fermions, respectively.
A straightforward way to calculate $G^K_{dL(R)}$ is to invert the matrix of the action defined in \eqref{eq:partition},
which becomes an efficient algorithm for numerical analyses of the average 
currents~\cite{PhysRevLett.92.127001,PhysRevB.71.024517,husmann2015,PhysRevA.98.041601,visuri2022}
With this method, however, it is not so easy to establish analytic structures of the currents.
To overcome this, we therefore adopt an approach based on Dyson equations~\cite{rammer2007}, where the action is split into non-perturbative and perturbative parts.
For the dissipative (three-terminal) situation, one can treat $S_{L(R)}$, the on-site energy term in $S_{1D}$, and $S_{\text{loss}}(S_3)$ as the former part,
and the hopping term in $S_{1D}$ and $S_T(\tilde{S}_T)$ as the latter part.
Under this setting, calculation in the presence of dissipation can be done similarly with one in the absence of dissipation.
We note that the analysis with Dyson equations becomes exact by treating the perturbative part at all order level, in which case
the result corresponds to one with the matrix inversion of the action.

\section{Single-site case}
In order to see the essential structure of the average particle and energy currents both in dissipative two-terminal and three-terminal systems,
 we start to analyze $L=1$ case, i.e., noninteracting quantum dot system. 
For application of Dyson equations in the Keldysh formalism in our system,
it is convenient to use the following relation~\cite{haug2008}:
\beq
\sum_{\mathbf{k}}t_{\mathbf{k}}G^K_{dL(R)}(\mathbf{k},\omega)=-\sum_{\mathbf{k}}|t_{\mathbf{k}}|^2[G^R_{d}(\omega)g^K_{L(R)}(\mathbf{k},\omega)\nonumber\\
+G^K_{d}(\omega)g^A_{L(R)}(\mathbf{k},\omega)],
\label{eq:langreth}
\eeq
where $g$ is uncoupled Green's function that is obtained from inverse of the so-called RAK matrix \eqref{eq:uncoupled},
and $G_{d}$ is full Green's function of the 1D chain that contains effects of the perturbative part.
Notice also that in order to obtain above, we applied the Langreth rule in the Keldysh formalism~\cite{rammer2007} .
Owing to this deformation, the remaining task is to calculate $G^{R/K}_{d}$.

\subsection{Dissipative situation}
We first look at the dissipative situation corresponding to the situation of Fig.~1~(a).
There, retarded and advanced Green's function in the single site obey the following Dyson equation~\cite{rammer2007}:
\beq
[G^{R(A)}_{d} (\omega) ]^{-1}= [g^{R(A)}_{d}(\omega) ]^{-1}-\Sigma^{R(A)}(\omega).
\eeq
We note that the retarded or advanced component is decoupled and therefore satisfies
$[G^{-1}]^{R(A)}=[G^{R(A)}]^{-1}$.
We also introduce retarded and advanced components of  the self-energy
\beq
\Sigma^{R(A)}(\omega)&&=\sum_{\mathbf{k}}|t_{\mathbf{k}}|^2[g^{R(A)}_{L}(\mathbf{k},\omega-\Delta\mu/2)\nonumber\\
&&\ \ +g^{R(A)}_R(\mathbf{k},\omega+\Delta\mu/2)].
\eeq
By introducing
\beq
R(\omega)&&\equiv2\sum_{\mathbf{k}}|t_{\mathbf{k}}|^2\text{Re}[g^{R}_{L}(\mathbf{k},\omega-\Delta\mu/2)]\nonumber\\
&&=2\sum_{\mathbf{k}}|t_{\mathbf{k}}|^2\text{Re}[g^{R}_{R}(\mathbf{k},\omega+\Delta\mu/2)],\\
\Gamma(\omega)&&\equiv-2\sum_{\mathbf{k}}|t_{\mathbf{k}}|^2\text{Im}[g^{R}_{L}(\mathbf{k},\omega-\Delta\mu/2)]\nonumber\\
&&=-2\sum_{\mathbf{k}}|t_{\mathbf{k}}|^2\text{Im}[g^{R}_{R}(\mathbf{k},\omega+\Delta\mu/2)],
\eeq
the full retarded and advanced Green's functions are obtained as
\beq
G^{R/A}_{d}(\omega)=\frac{1}{\omega-\epsilon_0-R(\omega) \pm i[\frac{\gamma}{2}+\Gamma(\omega)]}.
\label{eq:dissipative-gra}
\eeq
In addition, the Keldysh component obeys the following Dyson equation:
\beq
G^K_{d}(\omega)&&=[1+G^R_d(\omega)\Sigma^R(\omega)]g^K_d(\omega)[1+\Sigma^A(\omega)G^A_d(\omega)]\nonumber\\
&&\ \ +G^R_d(\omega)\Sigma^K(\omega)G^A_d(\omega).
\eeq
By using 
\beq
\Sigma^K(\omega)&&=\sum_{\mathbf{k}}|t_{\mathbf{k}}|^2[g^K_L(\mathbf{k},\omega-\Delta\mu/2)+g^K_R(\mathbf{k},\omega+\Delta\mu/2)]\nonumber\\
&&=-2i\Gamma(\omega)[1\pm n_L(\omega-\Delta\mu/2)\pm n_R(\omega+\Delta\mu/2)],\nonumber\\
\eeq
and
\beq
&&[1+G^R_d(\omega)\Sigma^R(\omega)]g^K_d(\omega)[1+\Sigma^A(\omega)G^A_d(\omega)]\nonumber\\
&&=G^R_d(\omega)[g^R_d(\omega)]^{-1}g^K_d(\omega)[g^A_d(\omega)]^{-1}G^A_d(\omega)\nonumber\\
&&=-G^R_d(\omega)[g^{-1}_d(\omega)]^KG^A_d(\omega),
\eeq
the Dyson equation for the Keldysh component can be solved as
\beq
&&G^K_{d}(\omega)\nonumber\\
&&=\frac{-2i[\frac{\gamma}{2}+\Gamma(\omega)]\mp2i\Gamma(\omega)[n_L(\omega-\Delta\mu/2)+ n_R(\omega+\Delta\mu/2)] }{ [\omega-\epsilon_0-R(\omega)]^2 
+[\frac{\gamma}{2}+\Gamma(\omega)]^2 }.\nonumber\\
\label{eq:dissipative-gk}
\eeq
In total, the average current is determined as
\beq
 I&&=\int_{-\infty}^{\infty}\frac{d\omega}{2\pi}\Big[ {\cal T}(\omega)+\frac{{\cal L}(\omega)}{2}\Big]  \nonumber\\
&& \ \ \times  [n_L(\omega-\Delta\mu/2)-n_R(\omega+\Delta\mu/2)],\label{eq:current-1}
\label{eq:average-1}
\eeq
where we introduce
\beq
{\cal T}(\omega)=\frac{[\Gamma(\omega)]^2}{(\omega-\epsilon_0-R(\omega))^2 +[\frac{\gamma}{2}+\Gamma(\omega)]^2},\\
{\cal L}(\omega)=\frac{\Gamma(\omega)\gamma}{(\omega-\epsilon_0-R(\omega))^2 +[\frac{\gamma}{2}+\Gamma(\omega)]^2}.
\eeq
Physically, ${\cal T}(\omega)$ and ${\cal L}(\omega)$ represent transmittance and loss probability, respectively.
When $\gamma=0$, ${\cal T}(\omega)$ is indeed reduced to transmittance in the noninteracting quantum dot system~\cite{PhysRevLett.68.2512,haug2008}.
In addition, when $\gamma\ne0$, ${\cal T}(\omega)$ is reduced to one in a phenomenological model with inelastic widths
in which ${\cal L}(\omega)$ is absent~\cite{jonson1987}.
On the other hand, the presence of ${\cal L}(\omega)$ is peculiar to the lossy system.
Physical meaning of ${\cal L}(\omega)$ becomes clearer by noting that
the average of the particle loss rate is expressed as
\beq
-\dot{N}&&=-\dot{N}_L-\dot{N}_R\nonumber\\
&&=\int_{-\infty}^{\infty}\frac{d\omega}{2\pi}{\cal L}(\omega)[n_L(\omega-\Delta\mu/2)+n_R(\omega+\Delta\mu/2)].\nonumber\\
\label{eq:loss-1}
\eeq
It is also instructive to point out the relation with Refs.~\cite{visuri2022,PhysRevB.102.205131}.
When the frequency dependence of $\Gamma(\omega)$ is neglected and $R(\omega)=0$, 
the particle current obtained in this work is identical to one in Refs.~\cite{visuri2022,PhysRevB.102.205131}.
The first condition is called wide-band approximation and is reasonable if low-energy transport such that the frequency dependence in $\Gamma$ is negligible is concerned.
In addition, the second condition means that the energy shift due to the couplings with the reservoirs is negligible, and
corresponds to the situation that the Lamb shift is neglected in the context of quantum optics.
For instance, these conditions are simultaneously satisfied 
if $t_{\mathbf{k}}$ is momentum independent such as the point-contact tunneling~\cite{PhysRevB.84.155414,PhysRevLett.118.105303}
and the constant density-of-states approximation in the reservoir that is often used in fermionic systems is reasonable.
In contrast, our formulation is available even if such conditions are not valid, which contains superfluid bosonic systems~\cite{PhysRevResearch.2.023284,PhysRevResearch.3.043058,PhysRevA.106.L011303}.

With the use of the results above,
we can also obtain the energy current.
To this end, we note
\beq
&&\sum_{\mathbf{k}}|t_{\mathbf{k}}|^2\epsilon_{\mathbf{k}}\text{Im}[g^R_{L/R}(\mathbf{k},\omega\mp\Delta\mu/2) ]\nonumber\\
&&=\sum_{\mathbf{k}}|t_{\mathbf{k}}|^2(\omega+\mu) \text{Im}[g^R_{L/R}(\mathbf{k},\omega\mp\Delta\mu/2) ],
\eeq
where we use $\text{Im}[g^R_{L/R}]\propto\delta(\omega+\mu-\frac{k^2}{2m})$ with the average chemical potential between reservoirs $\mu=\frac{\mu_L+\mu_R}{2}$.
By using this relation, the average energy current is shown to be expressed as
\beq
I_E&&=\int_{-\infty}^{\infty}\frac{d\omega}{2\pi}(\omega+\mu)\Big[ {\cal T}(\omega)+\frac{{\cal L}(\omega)}{2}\Big] \nonumber\\
&& \ \ \times[n_L(\omega-\Delta\mu/2)-n_R(\omega+\Delta\mu/2)].
\label{eq:energy-1}
\eeq

In order to compare with the result in Ref.~\cite{PhysRevA.100.053605},
we bear in mind that while $\omega$ adopted in Ref.~\cite{PhysRevA.100.053605} corresponds to the absolute energy,
$\omega=0$ in our formulation corresponds to the energy at average chemical potential.
By taking into account this energy shift between two formulations,
it turns out that the formal expressions of Eqs.~\eqref{eq:average-1},~\eqref{eq:loss-1}~\eqref{eq:energy-1} are  consistent with ones found in
the non-hermitian Landauer-B\"{u}ttiker analysis~\cite{PhysRevA.100.053605}.

\subsection{Three-terminal situation}
We now consider the situation that atom loss is absent but instead
the noninteracting dot is attached to three terminals (Fig.~1~(b)).
In this case,  the retarded and advanced components of the self-energy are given by
\begin{flalign}
\Sigma^{R(A)}(\omega)=\sum_{\mathbf{k}}[|t_{\mathbf{k}}|^2g^{R(A)}_{L}(\mathbf{k},\omega-\Delta\mu/2)\nonumber\\
 +|t_{\mathbf{k}}|^2g^{R(A)}_R(\mathbf{k},\omega+\Delta\mu/2) +|\tilde{t}_{\mathbf{k}}|^2g^{R(A)}_3(\mathbf{k},\omega+V)],
\end{flalign}
where $V\equiv \mu-\mu_3$ with the chemical potential of the third reservoir $\mu_3$. As in the case of the previous subsection, we introduce
 \beq
 R_3(\omega)\equiv2\sum_{\mathbf{k}}|\tilde{t}_{\mathbf{k}}|^2\text{Re}[g^{R}_{3}(\mathbf{k},\omega+V)],\\
 \Gamma_3(\omega)\equiv-2\sum_{\mathbf{k}}|\tilde{t}_{\mathbf{k}}|^2\text{Im}[g^{R}_{3}(\mathbf{k},\omega+V)].
 \eeq
Then, full retarded and advanced Green's functions are obtained as
 \begin{flalign}
 G^{R/A}_{d}(\omega)=\frac{1}{\omega-\epsilon_0-R(\omega)-\frac{R_3(\omega)}{2} \pm i[\Gamma(\omega)+\frac{\Gamma_3(\omega)}{2}]}.
 \label{eq:three-gra}
 \end{flalign}
In addition, the Keldysh component of the self-energy is 
\beq
\Sigma^K(\omega)&&=-2i\Gamma(\omega)[1\pm n_L(\omega-\Delta\mu/2)\pm n_R(\omega+\Delta\mu/2)]\nonumber\\
&&-i\Gamma_3(\omega)[1\pm 2n_3(\omega+V)],
\eeq
where upper and lower signs are for bosons and fermions, respectively, and
$n_3$ is the distribution function of the third reservoir.
Since $[g^{-1}]^K=0$ in the absence of the single particle loss, we obtain
\beq
&&G^K_{d}(\omega)=\frac{-i\Gamma_3(\omega)[1\pm 2n_3(\omega+V)] }{ [\omega-\epsilon_0-R(\omega)+\frac{R_3(\omega)}{2}]^2 +
[\frac{\Gamma_3(\omega)}{2}-\Gamma(\omega)]^2 }\nonumber\\
&&-\frac{2i\Gamma(\omega)[1\pm n_L(\omega-\Delta\mu/2)\pm n_R(\omega+\Delta\mu/2)]}{ [\omega-\epsilon_0-R(\omega)-\frac{R_3(\omega)}{2}]^2 +[\frac{\Gamma_3(\omega)}{2}+\Gamma(\omega)]^2}.
\label{eq:three-gk}
\eeq

Now that full Green's functions are obtained, we turn to compare
Eqs.~\eqref{eq:dissipative-gra},~\eqref{eq:dissipative-gk} and Eqs.~\eqref{eq:three-gra}, \eqref{eq:three-gk}.
Then, it turns out that full Green's functions in the three terminal system coincide with ones in the dissipative two-terminal system when the following conditions are 
satisfied:
\beq
&&\Gamma_3(\omega)=\gamma,\label{eq:wide-band}\\
&&R_3(\omega)=0,\label{eq:rabishift}\\
&&n_3(\omega+V)=0.\label{eq:distri}
\eeq
Equations~\eqref{eq:wide-band} and~Eq.~\eqref{eq:rabishift} represents the wide-band approximation and disregard of the Lamb shift, respectively.
In addition, Eq.~\eqref{eq:distri} means that  the third reservoir is empty.
The last condition is physically sound, since the dissipator in Eq.~\eqref{eq:lindblad-loss} (terms proportional to $\gamma$)
does not contain the gain effect~\cite{breuer2002}.
In total, when Eqs.~\eqref{eq:wide-band},~\eqref{eq:rabishift}, and~\eqref{eq:distri} are satisfied,
it follows that the transport quantities calculated in the previous subsection such as particle and heat current and particle loss rate also coincide with those
in the lossy two-terminal system.

Although the conditions of~\eqref{eq:wide-band},~\eqref{eq:rabishift}, and \eqref{eq:distri}
are necessary for the exact correspondence including the frequency dependences of ${\cal T}$ and ${\cal L}$,
one can relax them as far as a formal correspondence is concerned.
In particular, it is straightforward to confirm that  without~\eqref{eq:wide-band},~\eqref{eq:rabishift}, and \eqref{eq:distri},
the average particle and energy currents from left to right are expressed as
\beq
 I&&=\int_{-\infty}^{\infty}\frac{d\omega}{2\pi}\Big[ \tilde{{\cal T}}(\omega)+\frac{\tilde{{\cal L}}(\omega)}{2}\Big]  \nonumber\\
&& \ \ \times  [n_L(\omega-\Delta\mu/2)-n_R(\omega+\Delta\mu/2)],\label{eq:single-particle-2}
\eeq
\beq
 I_E&&=\int_{-\infty}^{\infty}\frac{d\omega}{2\pi}(\omega+\mu)\Big[ \tilde{{\cal T}}(\omega)+\frac{\tilde{{\cal L}}(\omega)}{2}\Big]  \nonumber\\
&& \ \ \times  [n_L(\omega-\Delta\mu/2)-n_R(\omega+\Delta\mu/2)],\label{eq:single-energy-2}
\eeq
where 
\beq
\tilde{{\cal T}}(\omega)=\frac{[\Gamma(\omega)]^2}{(\omega-\epsilon_0-R(\omega)-\frac{R_3(\omega)}{2})^2 +[\frac{\Gamma_3(\omega)}{2}+\Gamma(\omega)]^2},
 \\
\tilde{{\cal L}}(\omega)=\frac{\Gamma(\omega)\Gamma_3(\omega)}{(\omega-\epsilon_0-R(\omega)-\frac{R_3(\omega)}{2})^2 +[\frac{\Gamma_3(\omega)}{2}+\Gamma(\omega)]^2}. 
\eeq
We note that Eqs.~\eqref{eq:single-particle-2} and~\eqref{eq:single-energy-2}
are formally equivalent to Eqs.~\eqref{eq:current-1} and~\eqref{eq:energy-1} 
in that the currents are expressed with transmittance and loss probability in a similar manner.
The only difference from the dissipative situation is that the additional frequency dependences 
originating from frequency dependence of $g_3$ and momentum dependence of $\tilde{t}_k$
come out in $\tilde{{\cal T}}$ and $\tilde{{\cal L}}$.
What is especially remarkable is that the average currents do not depend on the distribution of the third reservoir.

It should also be noted that all the average quantities in the three-terminal situation
are not independent of $n_3$.
In fact, the average particle loss rate in the three-terminal situation, which is nothing but the particle gain rate of the third terminal, is expressed as
\beq
-\dot{N}=&&\int_{-\infty}^{\infty}\frac{d\omega}{2\pi}\Big[ \tilde{{\cal L}}(\omega)   \{n_L(\omega-\Delta\mu/2)+n_R(\omega+\Delta\mu/2)\}\nonumber\\
&&\ \ -2 \tilde{{\cal L}}(\omega)n_3(\omega+V)\Big].
\eeq
The expression above is not formally consistent with Eq.~\eqref{eq:loss-1}
due to the presence of the last term proportional to $n_3$.
Thus, for generic $\tilde{{\cal L}}$, the condition~\eqref{eq:distri} is essential to consistency of the particle loss rate between dissipative and three-terminal situations.
At the same time, notice that
\beq
\int d\omega \tilde{{\cal L}}(\omega)n_3(\omega+V)=0,
\eeq
can also be achieved for a non-empty third reservoir.
Indeed, the above equality is satisfied, if  $\tilde{{\cal L}}(\omega)=0$ for $\omega$'s such that $n_3(\omega+V)\ne0$.
Moreover, $\mu_3<\mu_L,\mu_R$ is also necessary to obtain nontrivial effects of $\tilde{{\cal L}}$ in transport.
In this case, the gain effect from the third reservoir is neglected and  lossy two-terminal transport corresponds to lossless three-terminal one.
 
\section{multi-site case}
Although the essential structure of the correspondence between the lossy two-terminal and lossless three-terminal systems
already appears in the single-site problem, 
in order to convince ourself of it, we now look at the multi-site case.

To this end, we point out the following relations:
\begin{flalign}
\sum_{\mathbf{k}}t_{\mathbf{k}}G^K_{dL}(\mathbf{k},\omega)=-\sum_{\mathbf{k}}t^2_{\mathbf{k}}[G^R_{d,11}(\omega)g^K_{L}(\mathbf{k},\omega-\Delta\mu/2)\nonumber\\
+G^K_{d,11}(\omega)g^A_{L}(\mathbf{k},\omega-\Delta\mu/2) ],\\
\sum_{\mathbf{k}}t_{\mathbf{k}}G^K_{dR}(\mathbf{k},\omega)=-\sum_{\mathbf{k}}t^2_{\mathbf{k}}[G^R_{d,LL}(\omega)g^K_{R}(\mathbf{k},\omega+\Delta\mu/2)\nonumber\\
+G^K_{d,LL}(\omega)g^A_{R}(\mathbf{k},\omega+\Delta\mu/2) ],
\end{flalign}
which is the multi-site generalization of Eq.~\eqref{eq:langreth}.
Here, we introduce the following $L\times L$ matrix of Green's function in the 1D chain:
\beq
\mathbf{G}_d=\begin{pmatrix}
G_{d,11} & G_{d,12} & \cdots & & G_{d,1L} \\
G_{d,21} & G_{d,22} & & & \vdots \\ 
\vdots & &  \ddots &  &\vdots \\
G_{d,L1} & G_{d,L2} &\cdots &  & G_{d,LL}
\end{pmatrix}.
\eeq
For example, $G^R_{d,11(LL)}(\tau,\tau')=-i\theta(\tau-\tau')\langle[d_{-N(N)}(\tau),d^{\dagger}_{-N(N)}(\tau')]_{\pm} \rangle$.
As in the case of the single site, 
what we need to do for determination of current formulas is to evaluate the retarded, advanced, and Keldysh components of $\mathbf{G}_d$.

\subsection{Dissipative situation}
In the multi-site case, the Dyson equation for retarded and advanced components is expressed as
\beq
[\mathbf{G}^{R/A}_d(\omega)]^{-1}=[\mathbf{g}^{R/A}_d(\omega)]^{-1}-\mathbf{\Sigma}^{R/A}(\omega).
\eeq
Since
\begin{widetext}
\beq
[\mathbf{g}^{R/A}_d]^{-1}=\begin{pmatrix}
\omega-\epsilon_N\pm i0^+ & 0 & \cdots & & & 0  \\
0 & \omega-\epsilon_{N-1}\pm i0^+ & 0  & & & \\
 \vdots & 0 & \ddots \\
\vdots  &\vdots & & \omega-\epsilon_0\pm i\frac{\gamma}{2} &  \\
  &  &  & &\ddots \\
 0 & \cdots & & & & \omega-\epsilon_N\pm i0^+ & 
\end{pmatrix},
\eeq
and
\beq
\mathbf{\Sigma}^{R/A}=\begin{pmatrix}
\frac{R(\omega)\mp i\Gamma(\omega)}{2} & -t_{N-1} & 0 &\cdots  & & 0 \\
-t_{N-1} & 0 & -t_{N-2} &  & &\\
 0& -t_{N-2} & 0 &  & &\\ 
\vdots \\
  & & & &  0 & -t_{N-1}\\
0&\cdots & & & -t_{N-1} & \frac{R(\omega)\mp i\Gamma(\omega)}{2}
\end{pmatrix},
\eeq
we obtain
\beq
\mathbf{G}_d^{R/A}(\omega)&&=\begin{pmatrix}
\omega-\epsilon_N-\frac{R(\omega)}{2}\pm i\frac{\Gamma(\omega)}{2} & t_{N-1} & 0 & &\cdots & &  0 \\
t_{N-1} & \omega-\epsilon_{N-1}\pm i0^+ & t_{N-2} &  &  & & & & &\\
0 &  &\ddots & & & & \\
\vdots & \ & &   & \omega-\epsilon_0\pm i\frac{\gamma}{2} &  & \\
 & & & & & \ddots  & t_{N-1}\\
0 &  & & & &  t_{N-1} & \omega-\epsilon_N-\frac{R(\omega)}{2}\pm i\frac{\Gamma(\omega)}{2}
\end{pmatrix}^{-1}.
\label{eq:matrix-gad}
\eeq
Thus, it turns out that the inverse of retarded or advanced Green's function is expressed with the so-called tridiagonal matrix.
In general, a non-singular tridiagonal matrix
\beq
T=\begin{pmatrix}
a_1 & b_1 & 0 & 0\\
c_1 & a_2  & \ddots & 0\\
0 & \ddots & \ddots & b_{L-1}\\
0 & 0&  c_{L-1} & a_L
\end{pmatrix},
\eeq
is known to be inverted as~\cite{da2007,usmani1994,PhysRevB.104.144301,PhysRevResearch.4.013109},
\beq
T^{-1}_{i,j}=\begin{cases}
(-1)^{i+j}b_i\cdots b_{j-1}\theta_{i-1}\phi_{j+1}/\theta_L \ \ i<j\\
\theta_{i-1}\phi_{j+1}/\theta_L \ \ \ \ \ \ \ \ \ \ \ \ \ \ \ \ \ \ \ \ \ \  \  \ \ i=j\\
(-1)^{i+j}c_j\cdots c_{i-1}\theta_{j-1}\phi_{i+1}/\theta_L \ \ i>j
\end{cases}
\eeq
Here, $\theta_i$ and $\phi_i$ satisfy the following recurrence relations:
\beq
&&\theta_i=a_i\theta_{i-1}-b_{i-1}c_{i-1}\theta_{i-2}, \ \ i=2,3,\cdots,L\\
&&\phi_i=a_i\phi_{i+1}-b_ic_i\phi_{i+2}, \  \ \ \ \ \ \ i=L-1,\cdots,1
\eeq
with initial conditions $\theta_0=1$, $\theta_1=a_1$, $\phi_{L+1}=1$, and $\phi_L=a_L$.
Thus, by applying the above formula, $\mathbf{G}_d^{R/A}$ can exactly be obtained.

On the other hand, since
\beq
[\mathbf{g}^{-1}_{d}]^K=\begin{pmatrix}
0 & \cdots &  & & 0\\
\vdots &\ddots  \\
 & &  i\gamma & \\
 & & &\ddots \\
0 & \cdots & && 0
\end{pmatrix},
\eeq
and
\beq
\mathbf{\Sigma}^K=\begin{pmatrix}
-i\Gamma(\omega)[1\pm2n_L(\omega-\Delta\mu/2)] & 0 & \cdots & 0\\
0 & 0 \\
\vdots&  & \ddots \\
0 & \cdots & & -i\Gamma(\omega)[1\pm2n_R(\omega+\Delta\mu/2)]
\end{pmatrix},
\eeq
the Keldysh component is expressed as
\beq
\mathbf{G}^K_{d}(\omega)&&=-\mathbf{G}^R_d(\omega)[\mathbf{g}^{-1}_d(\omega)]^K\mathbf{G}^A_d(\omega)+\mathbf{G}^R_d(\omega)\mathbf{\Sigma}^K(\omega)
\mathbf{G}^A_d(\omega)\nonumber\\
&&=\mathbf{G}^R_d(\omega)\begin{pmatrix}
-i\Gamma(\omega)[1\pm2n_L(\omega-\Delta\mu/2)] & 0 & \cdots & & &  0 \\
0 & 0 \\
\vdots & & \ddots \\
 & & &  -i\gamma &  & \\
& & & & \ddots  \\
0 & \cdots & & & & -i\Gamma(\omega)[1\pm2n_R(\omega+\Delta\mu/2)]
\end{pmatrix}
\mathbf{G}^A_d(\omega).
\label{eq:matrix-kd}
\eeq

Based on the results obtained above, we obtain the average particle current and particle loss rate.
By using symmetries such as $G^{R/A}_{d11}=G^{R/A}_{dLL}$ and $G^{R/A}_{d,1L}=G^{R/A}_{d,L1}$,
the particle current is expressed as 
\beq
I=\int_{-\infty}^{\infty}\frac{d\omega}{2\pi}\Gamma(\omega)\Big[-\text{Im}[G^R_{d,11}(\omega)]
-\frac{\Gamma(\omega)\{|G^R_{d,11}(\omega)|^2-|G^R_{d,1L}(\omega)|^2\}}{2}\Big][n_L(\omega-\Delta\mu/2)-n_R(\omega+\Delta\mu/2) ].
\eeq
In addition, the particle loss rate is expressed as
\beq
-\dot{N}&&=\int_{-\infty}^{\infty}\frac{d\omega}{2\pi}\Big[\mp\Gamma(\omega) \{2\text{Im}[G^R_{d,11}(\omega)]
+\Gamma(\omega)\{ |G^R_{d,11}|^2+ |G^R_{d,1L}|^2\}+\gamma|G^R_{d,1\frac{L+1}{2}}(\omega)|^2
\}\nonumber\\
&&+\Gamma(\omega) \{-\Gamma(\omega)\{|G^R_{d,11}|^2+|G^R_{d,1L}|^2\}-2\text{Im}[G^R_{d,11}(\omega)]\}[n_L(\omega-\Delta\mu/2)+n_R(\omega+\Delta\mu/2)]\Big].
\eeq
In order to obtain convenient forms, we consider the following identity:
\beq
\mathbf{G}^R_d-\mathbf{G}^A_d=\mathbf{G}^R_d\{(\mathbf{G}_d^A)^{-1}-(\mathbf{G}_d^R)^{-1}\}\mathbf{G}^A_d.
\eeq
This leads to
\beq
2\text{Im}[G^R_{d,11}(\omega)]=-\Gamma(\omega)\{|G^R_{d,11}|^2+|G^R_{d,1L}|^2\}-\gamma|G^R_{d,1\frac{L+1}{2}}(\omega)|^2.
\eeq
Thus, the particle current and particle loss rate are obtained as
\beq
&&I=\int_{-\infty}^{\infty}\frac{d\omega}{2\pi}\Big[{\cal T}^L(\omega)+\frac{{\cal L}^L(\omega)}{2}\Big][n_L(\omega-\Delta\mu/2)-n_R(\omega+\Delta\mu/2)],
\label{eq:generic-current}\\
&&-\dot{N}=\int_{-\infty}^{\infty}\frac{d\omega}{2\pi}{\cal L}^L(\omega)[n_L(\omega-\Delta\mu/2)+n_R(\omega+\Delta\mu/2)],
\label{eq:generic-loss}
\eeq
where
\beq
&&{\cal T}^L(\omega)=[\Gamma(\omega)]^2|G^R_{d,1L}(\omega)^2|,\label{eq:generic-t}\\
&&{\cal L}^L(\omega)=\Gamma(\omega) \Big[-\Gamma(\omega)\{|G^R_{d,11}|^2+|G^R_{d,1L}|^2\}-2\text{Im}[G^R_{d,11}(\omega)]\Big],\label{eq:generic-l}
\eeq
are transmittance and loss probability, respectively. It is also straightforward to show that the energy current is obtained as
\beq
I_E=\int_{-\infty}^{\infty}\frac{d\omega}{2\pi}(\omega+\mu)\Big[{\cal T}^L(\omega)+\frac{{\cal L}^L(\omega)}{2}\Big][n_L(\omega-\Delta\mu/2)-n_R(\omega+\Delta\mu/2)].
\label{eq:generic-energy}
\eeq
Equations~\eqref{eq:generic-current}, \eqref{eq:generic-loss}, and \eqref{eq:generic-energy} are
multi-site generalization of Eqs.~\eqref{eq:current-1}, \eqref{eq:loss-1}, and~\eqref{eq:energy-1}.

In the case of $L=1$, it is easy to check that the transmittance (Eq.~\eqref{eq:generic-t}) and loss probability (Eq.~\eqref{eq:generic-l}) 
coincide with those appeared in the single-site case.
For multi sites, explicit forms of Eqs.~\eqref{eq:generic-t} and \eqref{eq:generic-l} in generic 1D chain are lengthly.
If $\epsilon\equiv\epsilon_0=\epsilon_1=\cdots \epsilon_N$ and $t\equiv t_0=t_1=\cdots t_{N-1}$, and $R(\omega)=0$,
 however, one can obtain simpler expressions.
For $L=3$ case, we obtain
\beq
&&{\cal T}^{L=3}(\omega)=\frac{[\Gamma(\omega)]^2t^4}{(\omega-\epsilon+[\Gamma(\omega)]^2/4)[ (\omega-\epsilon)^4+(\omega-\epsilon)^2([\Gamma(\omega)]^2/4-4t^2+\gamma^2/4)
+(2t^2+\Gamma\gamma/4)^2 ] },\\
&&{\cal L}^{L=3}(\omega)=\frac{\Gamma(\omega) t^2\gamma }{ (\omega-\epsilon)^4+(\omega-\epsilon)^2([\Gamma(\omega)]^2/4-4t^2+\gamma^2/4)
+(2t^2+\Gamma(\omega)\gamma/4)^2 },
\eeq
and for $L=5$ case, we obtain
\beq
&&{\cal T}^{L=5}(\omega)=[\Gamma(\omega)]^2t^8/
[(\omega-\epsilon)^2[\Gamma(\omega)]^2/4+((\omega-\epsilon)^2-t^2)^2]\nonumber\\
&&\times[ (\omega-\epsilon)^6
+t^4(\Gamma(\omega)+\gamma/2)^2+(\omega-\epsilon)^4([\Gamma(\omega)]^2/4-6t^2+\gamma^2/4)+(\omega-\epsilon)^2\{t^2(9t^2-\gamma^2/2)
+(-4t^2+\gamma^2/4)[\Gamma(\omega)]^2/4 \} ],\nonumber\\
 \eeq
 \beq
{\cal L}^{L=5}(\omega)=\Gamma(\omega) t^4\gamma/ \Big[(\omega-\epsilon)^6
+t^4(\Gamma(\omega)+\gamma/2)^2+(\omega-\epsilon)^4([\Gamma(\omega)]^2/4-6t^2+\gamma^2/4)\nonumber\\
+(\omega-\epsilon)^2\{t^2(9t^2-\gamma^2/2)
+(-4t^2+\gamma^2/4)[\Gamma(\omega)]^2/4 \} \Big].
\eeq
In this way, one can obtain the explicit forms of ${\cal T}^L(\omega)$ and ${\cal L}^L(\omega)$,
provided that the number of the sites is fixed and the corresponding $G^R_{d,11}$ and $G^R_{d,1L}$ are determined.

\subsection{Three-terminal situation}
In the case of the three terminals, the retarded or advanced component of the self-energy is given by
\beq
\mathbf{\Sigma}^{R/A}=\begin{pmatrix}
\frac{R(\omega)\mp i\Gamma(\omega)}{2} & -t_{N-1} & 0 &\cdots  & & 0 \\
-t_{N-1} & 0 & -t_{N-2} &  & &\\
 0& & \ddots &  & &\\ 
\vdots & & & \frac{R_3(\omega)\mp i\Gamma_3(\omega)}{2}  \\
  & & & &  \ddots &t_{N-1} \\
0&\cdots & & & -t_{N-1} & \frac{R(\omega)\mp i\Gamma(\omega)}{2}
\end{pmatrix},
\eeq
we obtain
\beq
\mathbf{G}_d^{R/A}&&=\begin{pmatrix}
\omega-\epsilon_N-\frac{R(\omega)}{2}\pm i\frac{\Gamma(\omega)}{2} & t_{N-1} & 0 & &\cdots & &  0 \\
t_{N-1} & \omega-\epsilon_{N-1}\pm i0^+ & t_{N-2} &  &  & & & & &\\
0 &  &\ddots & & & & \\
\vdots & \ & &   & \omega-\epsilon_0-\frac{R_3(\omega)}{2}\pm i\frac{\Gamma_3(\omega)}{2} &  & \\
 & & & & & \ddots &t_{N-1} \\
0 &  & & & &  t_{N-1} & \omega-\epsilon_N-\frac{R(\omega)}{2}\pm i\frac{\Gamma(\omega)}{2}
\end{pmatrix}^{-1}.\nonumber\\
\label{eq:matrix-ga3}
\eeq
In addition, the absence of the particle loss means $[\mathbf{g}^{-1}_d(\omega)]^K=\mathbf{0}$, and therefore, 
the Keldysh component of the self-energy is given by
\beq
\mathbf{\Sigma}^K=\begin{pmatrix}
-i\Gamma(\omega)[1\pm2n_L(\omega-\Delta\mu/2)] & 0 & \cdots & & & 0\\
0 & 0 \\
\vdots&  & \ddots \\
& & & -i\Gamma_3(\omega)[1\pm2n_3(\omega+V)] \\
& & & & \ddots \\
0 & \cdots & & & & -i\Gamma(\omega)[1\pm2n_R(\omega+\Delta\mu/2)]
\end{pmatrix}.
\eeq
Thus, the Keldysh component of full Green's function is obtained as
\beq
&&\mathbf{G}^K_{d}(\omega)=-\mathbf{G}^R_d(\omega)[\mathbf{g}^{-1}_d(\omega)]^K\mathbf{G}^A_d(\omega)+\mathbf{G}^R_d(\omega)\mathbf{\Sigma}^K(\omega)
\mathbf{G}^A_d(\omega)\nonumber\\
&&=\mathbf{G}^R_d(\omega)\begin{pmatrix}
-i\Gamma(\omega)[1\pm2n_L(\omega-\Delta\mu/2)] & 0 & \cdots & & &  0 \\
0 & 0 \\
\vdots & & \ddots \\
 & & &  -i\Gamma_3(\omega)[1\pm2n_3(\omega+V)] &  & \\
& & & & \ddots  \\
0 & \cdots & & & & -i\Gamma(\omega)[1\pm2n_R(\omega+\Delta\mu/2)]
\end{pmatrix}
\mathbf{G}^A_d(\omega).\nonumber\\
\label{eq:matrix-k3}
\eeq
By comparing Eqs.~\eqref{eq:matrix-ga3} and~\eqref{eq:matrix-k3}
with Eqs.~\eqref{eq:matrix-gad}  and~\eqref{eq:matrix-kd},
full Green's functions in the three-terminal system correspond to ones in the lossy two-terminal system
when Eqs.~\eqref{eq:wide-band}, \eqref{eq:rabishift}, and~\eqref{eq:distri} are satisfied.
In this case, it follows that the current formulas of the three terminal system obey
Eqs.~\eqref{eq:generic-current}, \eqref{eq:generic-loss}, and~\eqref{eq:generic-energy}.

When it comes to the formal correspondence between lossy and lossless transport,
we can relax those conditions as in the case of the single site.
Indeed, the average particle and energy currents are generally shown to be reduced to
Eqs.~\eqref{eq:generic-current} and~\eqref{eq:generic-energy}, respectively.
In addition, 
by using the following relation:
\beq
2\text{Im}[G^R_{d,11}(\omega)]=-\Gamma(\omega)\{|G^R_{d,11}(\omega)|^2+|G^R_{d,1L}(\omega)|^2 \}-\Gamma_3(\omega)|G^R_{d,1\frac{L+1}{2}}(\omega)|^2,
\eeq
the particle loss rate is obtained as
\beq
-\dot{N}=\int_{-\infty}^{\infty}\frac{d\omega}{2\pi}{\cal L}^L(\omega)[n_L(\omega-\Delta\mu/2)+n_R(\omega+\Delta\mu/2)-2n_3(\omega+V)],
\eeq
which is consistent with the single-site case.

\end{widetext}

\section{Three-terminal Landauer-B\"{u}ttiker analysis}
So far, we have examined the correspondence between the lossy two-terminal and lossless three-terminal systems.
When this correspondence holds, the current formulas are given by Eqs.~\eqref{eq:generic-current}, \eqref{eq:generic-loss}, and~\eqref{eq:generic-energy}
whose formal expressions are consistent with the non-hermitian Landauer-B\"{u}ttiker analysis.
Such a coincidence anticipates that the similar result may be obtained with the three-terminal Landauer-B\"{u}ttiker analysis (Fig.~1~(c)).

To see this, we point out that in the Landauer-B\"{u}ttiker formalism, the particle current operator in each reservoir is given by
\beq
\hat{I}_j(\tau)=\int\frac{d\omega}{2\pi}\int d\omega' e^{i(\omega-\omega')\tau}[\hat{a}_j^{\dagger}(\omega)\hat{a}_j(\omega')-\hat{b}^{\dagger}_j(\omega)\hat{b}_j(\omega)],\nonumber\\
\eeq
where $j=L,R,3$, and $\hat{a}$ and $\hat{b}$ are annihilation operators for incoming and outgoing states, respectively.
We note that the positive current direction is from each reservoir to channel.
The operators $\hat{a}$ and $\hat{b}$ are connected through the following symmetric $S$ matrix:
\beq
\begin{pmatrix}
b_L\\
b_R\\
b_3
\end{pmatrix}=
\begin{pmatrix}
r & t & t_L\\
t' & r' & t_R\\
t_L' & t_R' & r_3
\end{pmatrix}
\begin{pmatrix}
a_L\\
a_R\\
a_3
\end{pmatrix}.
\eeq
In the Landauer-B\"{u}ttiker, it is assumed that the incoming operators ($\hat{a}_j$ and $\hat{a}^{\dagger}_j$) obey
\beq
&&\langle \hat{a}^{\dagger}_L(\omega)\hat{a}_{L}(\omega')\rangle=\delta(\omega-\omega')n_L(\omega-\Delta\mu/2),\\
&&\langle \hat{a}^{\dagger}_R(\omega)\hat{a}_{R}(\omega')\rangle=\delta(\omega-\omega')n_R(\omega+\Delta\mu/2),\\
&&\langle \hat{a}^{\dagger}_3(\omega)\hat{a}_{3}(\omega')\rangle=\delta(\omega-\omega')n_3(\omega+V),\\
&&\langle \hat{a}^{\dagger}_i(\omega)\hat{a}_{j}(\omega')\rangle=0 \ \ i\ne j.
\eeq
In addition, the $S$ matrix obeys the unitary condition
\beq
SS^{\dagger}=1,
\eeq
which leads to
\beq
|r|^2+|t|^2+|t_L|^2=1,\\
|r'|^2+|t'|^2+|t_R|^2=1,\\
|r_3|^2+|t_L'|^2+|t_R'|^2=1.
\eeq
Thus, each average current is obtained as
\beq
I_L&&=\int_{-\infty}^{\infty}\frac{d\omega}{2\pi}[|t(\omega)|^2[n_L(\omega-\Delta\mu/2)-n_R(\omega+\Delta\mu/2)]\nonumber\\
&&\ \ +|t_L(\omega)|^2[n_L(\omega-\Delta\mu/2)-n_3(\omega+V)]],
\eeq
\beq
 I_R&&=\int_{-\infty}^{\infty}\frac{d\omega}{2\pi}[|t'(\omega)|^2[n_R(\omega+\Delta\mu/2)-n_L(\omega-\Delta\mu/2)]\nonumber\\
 &&\ \ +|t_R(\omega)|^2[n_R(\omega+\Delta\mu/2)-n_3(\omega+V)]],
 \eeq
 \beq
 I_3&&=\int_{\infty}^{\infty}\frac{d\omega}{2\pi}[|t_L'(\omega)|^2[n_3(\omega+V)-n_L(\omega-\Delta\mu/2)]\nonumber\\
&&\ \  +|t_R'(\omega)|^2[n_3(\omega+V)-n_R(\omega+\Delta\mu/2)]].
\eeq
Uo to now, we were dedicated to the generic three-terminal Landauer-B\"{u}ttiker analysis.
In order to obtain a connection to the dissipative two-terminal system discussed in the previous sections, we consider 
\beq
&&|t|^2=|t'|^2, \\
&&|t_L|^2=|t_R|^2,
\eeq
The first condition above represents the symmetry of the conduction channel such that
the transmittance from left to right reservoirs is identical to one from right to left reservoirs.
In addition, the second condition means that
the transmittance into the third reservoir is common between left and right reservoirs.
For example, an $S$-matrix introduced in Ref.~\cite{PhysRevA.30.1982} satisfies the these conditions.

When the above conditions hold, the average particle current from left to right is reduced to
\beq
I&&=\frac{I_L-I_R }{2}
=\int_{-\infty}^{\infty}\frac{d\omega}{2\pi}\Big[|t(\omega)|^2+\frac{|t_L(\omega)|^2}{2}\Big]\nonumber\\
&&\ \ \times[n_L(\omega-\Delta\mu/2)-n_R(\omega+\Delta\mu/2)].
\eeq
In addition, the particle loss rate is
\beq
&&-\dot{N}=I_L+I_R\nonumber\\
&&=\int_{-\infty}^{\infty}\frac{d\omega}{2\pi}|t_L(\omega)|^2[n_L(\omega-\Delta\mu/2)+n_R(\omega+\Delta\mu/2)\nonumber\\
&&\ \ \ \ \ \ \ \ \ \ \ \ \ \ \ \ \ \ \  \ \ \ \ \  -2n_3(\omega+V)].
\eeq
Thus, by treating
\beq
&&|t_L(\omega)|^2={\cal L}(\omega),\\
&&|t(\omega)|^2={\cal T}(\omega),
\eeq
the expressions above are consistent with  ones obtained in the previous sections, and
\beq
\int d\omega {\cal L}(\omega)n_3(\omega)=0
\eeq
is necessary to neglect the gain effect from the third reservoir.
Moreover, we can obtain the following relation:
\beq
{\cal T}(\omega)+{\cal R}(\omega)+{\cal L}(\omega)=1,
\eeq
with reflectance ${\cal R}=|r|^2=|r'|^2$.
The relation above is exactly one obtained in Refs.~\cite{PhysRevA.100.053605,PhysRevLett.122.040402,PhysRevB.101.144301}.
We also note that 
by using the  following energy current operator:
\beq
\hat{I}_{E,j}(\tau)&&=
\int\frac{d\omega}{2\pi}\int d\omega' e^{i(\omega-\omega')\tau}\Big[\frac{\omega+\omega'}{2}+\mu\Big]\nonumber\\
&&\ \ \times[\hat{a}_j^{\dagger}(\omega)\hat{a}_j(\omega')-\hat{b}^{\dagger}_j(\omega)\hat{b}_j(\omega)],
\eeq
it is straightforward to see that the similar correspondence occurs in the energy 
current~\footnote{As pointed out in Sec.~III~A,  $\omega$ used in this paper is measured from the average chemical potential between left and right reservoirs.
In order to obtain the typical forms adopted in the Landauer-Buttiker analyses where $\omega$ corresponds to the absolute energy,
the energy shift between different $\omega$'s must be taken into account.}.

Although one may think that the non-hermitian Landauer-B\"{u}ttiker analysis may contain deficiency on the treatment of the quantum jump term,
it is indeed consistent with the three terminal analysis that does not drop any term.
The reason is interpreted as follows.
In the Landauer-B\"{u}ttiker formalism, scattering region and leads (reservoirs) play separate roles
and especially what happens in the scattering region is assumed  not to affect thermodynamics of the leads.
A modification of the distribution in a system occurs by the quantum jump term yet is irrelevant to the present Landauer-B\"{u}ttiker analysis,
since the dissipation takes place in the scattering region and
the distribution function appearing in the Landauer-B\"{u}ttiker formula is entirely determined by the leads.

\section{Summary and outlook}
In this work, we have investigated the analytic structure of the two-terminal flows through
the lossy one-dimensional chain. 
By using the analysis in the basis of the Dyson equations, we have succeeded in
obtaining the generic expressions for the particle current~\eqref{eq:generic-current}, the loss-rate~\eqref{eq:generic-loss}, and the energy current~\eqref{eq:generic-energy},
which are identical to ones in certain three-terminal system without particle loss.
Although the theoretical backgrounds are not exactly same between analyses with and without the Lindblad formulation,
the formal correspondence occurs if  the gain effect of the third reservoir is negligible.

It is also remarkable that the formal expressions on the currents 
are consistent with the non-hermitian Landauer-B\"{u}ttiker analysis~\cite{PhysRevA.100.053605} and with certain three-terminal Landauer-B\"{u}ttiker analysis. 
In the context of mesoscopic transport, incoherent scattering of electrons has been explained with introduction of additional reservoirs~\cite{PhysRevB.33.3020}.
Our analyses gave prescription and theoretical basis of lossy mesoscopic systems with the Landauer-B\"{u}ttiker formalism,
and pave the way to simulate lossy two-terminal transport with three-terminal transport and vice versa.

The universality discussed in this work may be applied for other mesoscopic systems.
In particular, applications both for bosonic~\cite{PhysRevResearch.2.023284,PhysRevResearch.3.043058,PhysRevA.106.L011303} 
and fermionic~\cite{husmann2015,PhysRevA.98.041601,PhysRevResearch.2.023340} superfluid reservoirs are relevant to cold-atom experiments.
It is also interesting to look into dissipation effects of mesoscopic spin transport  in which interactions neglected in this work may play important 
roles~\cite{krinner2016,PhysRevLett.117.255302,PhysRevLett.123.193605,zhang2019,PhysRevLett.123.180402,damanet2019,PhysRevResearch.2.023152}.
In addition, microscopic understanding of dissipation effects for internal mesoscopic transport systems~\cite{ono2021} could also be an interesting future work.

\section*{acknowledgment}
The author thanks T. Esslinger, P. Fabritius, T. Giamarchi, M.-Z. Huang, J. Mohan, M. Talebi, A.-M. Visuri, and S. Wili
for a bunch of stimulating discussions that motivate this work.
This work is supported by MEXT Leading Initiative for Excellent Young Researchers,
JSPS KAKENHI Grant No.~JP21K03436, and Matsuo Foundation.

\appendix

\section*{\appendixname\,: Derivation of the partition function}
In order  to make this paper self-contained,
here we give a derivation of the functional integral form of the partition function whose density matrix obeys the Lindblad master equation.
Our derivation relies on Ref.~\cite{sieberer2016}. where the bosonic action has explicitly been derived.

For this purpose, it is useful to consider the coherent state obeying
\beq
\hat{\psi}|\psi\rangle=\psi|\psi\rangle,\ \\
\langle\psi |\hat{\psi}^{\dagger}=\langle\psi|\bar{\psi},
\eeq
where $\hat{\psi}$ and $\hat{\psi}^{\dagger}$ are the field annihilation and creation operators, respectively,
and  $\psi $ and $\bar{\psi} $ are complex numbers for bosons
and are Grassmann numbers for fermions.
The coherent state satisfies the following over-complete relation:
\beq
\hat{1}=\int d[\bar{\psi},\psi]e^{-\bar{\psi}\psi}|\psi\rangle\langle\psi|,
\eeq
where $d[\bar{\psi},\psi]=\frac{d\bar{\psi} d\psi}{\pi}$ for bosons
and $d[\bar{\psi},\psi]=d\bar{\psi}d\psi$ for fermions.
In addition, the overlap of two coherent states is given by
\beq
\langle\psi|\psi' \rangle=e^{\bar{\psi}\psi'}.
\eeq
By means of the coherent state, the partition function $Z=\text{Tr}\rho(\tau)$ 
can be expressed as
\beq
Z=\int d[\bar{\psi},\psi]e^{-\bar{\psi}\psi}\langle\pm\psi|\hat{\rho}(\tau)|\psi\rangle.
\label{eq:z-coherent}
\eeq
Here, upper and lower signs are respectively  for bosons and fermions,
and especially the minus sign for fermions originates from the anti-commutation property of 
 Grassmann numbers~\cite{kamenev2011}.

We next look at the density matrix operator whose dynamics obeys the following Lindblad master equation:
\beq
\partial_{\tau}\hat{\rho}&&={\cal L}\hat{\rho}\nonumber\\
&&=-i[\hat{H},\hat{\rho}]+\gamma\Big[\hat{L}\rho \hat{L}^{\dagger}-\frac{\{\hat{L}^{\dagger}\hat{L},\hat{\rho}\}}{2}\Big],
\eeq
where ${\cal L}$ and $L$ are Liouvillian and Lindblad operator, respectively.
The solution of the above equation  is formally expressed as
\beq
\hat{\rho}(\tau_f)&&=e^{(\tau_f-\tau_0){\cal L}}\hat{\rho}(\tau_0)\nonumber\\
&&=\lim_{N\to\infty}(1+\delta_{\tau}{\cal L})^N\hat{\rho}(\tau_0).
\eeq
We now decompose the time evolution from $\tau_0$ to $\tau_f$ into a sequence of small steps of duration $\delta_{\tau}=(\tau_f-\tau_0)/N$, and
denotes the density matrix operator after the $n$th step at $\tau_n=\tau_0+\delta_{\tau}n$ by $\hat{\rho}_n=\hat{\rho}(\tau_n)$.
We then have
\beq
\hat{\rho}_{n+1}=(1+\delta_{\tau}{\cal L})\hat{\rho}_n+O(\delta_{\tau}^2).
\eeq
\begin{widetext}
By using the coherent state, the density matrix operator is expressed as
\beq
&&\hat{\rho}_n=\int d[\bar{\psi}^+_{n},\psi^+_{n},\bar{\psi}^-_{n},\psi^-_{n}] e^{-\bar{\psi}^+_{n}\psi^+_{n}-\bar{\psi}^-_{n}\psi^-_{n}  } |\pm\psi^+_{n} \rangle
\langle\pm\psi^+_{n}|\hat{\rho}_n|\psi^+_{n}\rangle\langle\psi^-_{n}|,\\
&&\hat{\rho}_{n+1}=\int  d[\bar{\psi}^+_{n+1},\psi^+_{n+1},\bar{\psi}^-_{n+1},\psi^-_{n+1}] e^{-\bar{\psi}^+_{n+1}\psi^+_{n+1}-\bar{\psi}^-_{n+1}\psi^-_{n+1}  }
 |\pm\psi^+_{n+1} \rangle\langle\pm\psi^+_{n+1}|\hat{\rho}_{n+1}|\psi^-_{n+1}\rangle\langle\psi^-_{n+1}|.
\eeq
Since
\beq
\langle\pm\psi^+_{n+1}|\hat{\rho}_{n+1}|\psi^-_{n+1}\rangle =\langle\pm\psi^+_{n+1}|\hat{\rho}_{n}|\psi^-_{n+1}\rangle 
+\delta_{\tau}\langle\pm\psi^+_{n+1}|{\cal L} \hat{\rho}_{n}|\psi^-_{n+1}\rangle,
\eeq
we have
\beq
\langle\pm\psi^+_{n+1}|\hat{\rho}_{n}|\psi^-_{n+1}\rangle=
\int  d[\bar{\psi}^+_{n},\psi^+_{n},\bar{\psi}^-_{n},\psi^-_{n}] 
e^{(\bar{\psi}^+_{n+1}-\bar{\psi}^+_{n})\psi^+_{n}+\bar{\psi}^-_{n}(\psi^-_{n+1} -\psi^-_{n} )}
\langle\pm\psi^+_{n}|\hat{\rho}_n|\psi^-_{n}\rangle,\nonumber\\
\eeq
\beq
&&\langle\pm\psi^+_{n+1}|{\cal L} \hat{\rho}_{n}|\psi^-_{n+1}\rangle=-i\langle\pm\psi^+_{n+1}| [\hat{H},\hat{\rho}_n] |\psi^-_{n+1}\rangle
+\gamma\langle\pm\psi^+_{n+1}|\Big(\hat{L}\hat{\rho}_n\hat{L}^{\dagger}-\frac{\{{\hat{L}^{\dagger}\hat{L},\hat{\rho}_n \}}}{2} \Big) |\psi^-_{n+1}\rangle\nonumber\\
&&=\int  d[\bar{\psi}^+_{n},\psi^+_{n},\bar{\psi}^-_{n},\psi^-_{n}] 
e^{(\bar{\psi}^+_{n+1}-\bar{\psi}^+_{n})\psi^+_{n}+\bar{\psi}^-_{n}(\psi^-_{n+1} -\psi^-_{n} )}
\nonumber\\
&&\Big[-iH^+_n+iH^-_n+\gamma\Big(L^+_n\bar{L}^-_n -\frac{\bar{L}^+_n L^+_n+\bar{L}^-_n L^-_n}{2} \Big) \Big]
\langle\pm\psi^+_{n}|\hat{\rho}_n|\psi^-_{n}\rangle.
\eeq
Here, $H^{+(-)}_n=H(\pm\bar{\psi}^{+(-)}_{n+1(n)},\pm\psi^{+(-)}_{n(n+1)})$
consist of fields on the $+(-)$ contour only, and the same holds true for $L_n^{\pm}$.
By using
\beq
\partial_{\tau}\psi^{\pm}_{n}=\frac{\psi^{\pm}_{n+1}-\psi^{\pm}_{n} }{\delta_{\tau}},
\eeq
we obtain
\beq
e^{(\bar{\psi}^+_{n+1}-\bar{\psi}^+_{n})\psi^+_{n}+\bar{\psi}^-_{n}(\psi^-_{n+1} -\psi^-_{n} )}
=e^{\delta_{\tau}[\partial_{\tau}\bar{\psi}^+_{n}\psi^+_{n} +\bar{\psi}^-_{n}\partial_{\tau}\psi^-_{n} ]}.
\eeq
In addition, we note
\beq
1+\delta_{\tau}\Big[-iH^+_n+iH^-_n+\gamma\Big(L^+_n\bar{L}^-_n -\frac{\bar{L}^+_nL^+_n+\bar{L}^-_n L^-_n}{2} \Big) \Big]
\approx e^{\delta_{\tau}\Big[-iH^+_n+iH^-_n+\gamma\Big(L^+_n\bar{L}^-_n -\frac{\bar{L}^+_n L^+_n+\bar{L}^-_n L^-_n}{2} \Big) \Big]}.
\eeq
Therefore, we obtain
\beq
\langle\pm\psi^+_{n+1}|\hat{\rho}_{n+1}|\psi^-_{n+1}\rangle=
\int  d[\bar{\psi}^+_{n},\psi^+_{n},\bar{\psi}^-_{n},\psi^-_{n}] 
 e^{\delta S_n}
\langle\pm\psi^+_{n}|\hat{\rho}_n|\psi^-_{n}\rangle,
\eeq
where
\beq
\delta S_n=\delta_{\tau}\Big[\bar{\psi}^+_ni\partial_{\tau}\psi^+_n-\bar{\psi}^-_ni\partial_{\tau}\psi^-_n
-H^+_n+H^-_n-i\gamma\Big(L^+_n\bar{L}^-_n -\frac{\bar{L}^+_n L^+_n+\bar{L}^-_n L^-_n}{2} \Big) \Big].
\eeq
\end{widetext}
Thus, the partition function is obtained as
\beq
Z&&=\text{Tr} e^{(\tau_f-\tau_0){\cal L}}\hat{\rho}(\tau_0)\nonumber\\
&&=\int {\cal D}[\bar{\psi}^+,\psi^+,\bar{\psi}^-,\psi^-]e^{iS}\langle\pm\psi^+(\tau_0)|\hat{\rho}(\tau_0)|\psi^-(\tau_0) \rangle.\nonumber\\
\eeq
When we are interested in  a steady state, we take $\tau_0\to-\infty$ and $\tau_f\to\infty$.
There, we can make an assumption that the initial state in the infinite past does not affect the steady state, and 
we can ignore the boundary term, $\langle\pm\psi^+(\tau_0)|\hat{\rho}(\tau_0)|\psi^{-}(\tau_0) \rangle$.
In this case, the partition function is reduced to
\beq
Z=\int {\cal D}[\bar{\psi}^+,\psi^+,\bar{\psi}^-,\psi^-]e^{iS}.
\eeq

Finally, we apply the above formula into our system, where $\hat{L}=\hat{d}_0$.
There, we should beware of the sign of the  quantum jump term in the action $-i\gamma L^{+}\bar{L}^-$.
Especially, for fermions where the anti-periodic boundary condition is adopted~\eqref{eq:z-coherent},
the jump term is expressed as $i\gamma d_0\bar{d}_0$, which has the opposite sign to bosons.
This is in contrast to systems with two-particle loss where the sign different between fermions and bosons does not matter,
since the Lindblad operator is bosonic~\cite{PhysRevLett.127.055301}. 
By using the anti-commutation relation for  Grassmann numbers, however, we obtain 
the partition function~\eqref{eq:partition} whose formal expression is independent of quantum statistics.


\begin{thebibliography}{62}%
\makeatletter
\providecommand \@ifxundefined [1]{%
 \@ifx{#1\undefined}
}%
\providecommand \@ifnum [1]{%
 \ifnum #1\expandafter \@firstoftwo
 \else \expandafter \@secondoftwo
 \fi
}%
\providecommand \@ifx [1]{%
 \ifx #1\expandafter \@firstoftwo
 \else \expandafter \@secondoftwo
 \fi
}%
\providecommand \natexlab [1]{#1}%
\providecommand \enquote  [1]{``#1''}%
\providecommand \bibnamefont  [1]{#1}%
\providecommand \bibfnamefont [1]{#1}%
\providecommand \citenamefont [1]{#1}%
\providecommand \href@noop [0]{\@secondoftwo}%
\providecommand \href [0]{\begingroup \@sanitize@url \@href}%
\providecommand \@href[1]{\@@startlink{#1}\@@href}%
\providecommand \@@href[1]{\endgroup#1\@@endlink}%
\providecommand \@sanitize@url [0]{\catcode `\\12\catcode `\$12\catcode
  `\&12\catcode `\#12\catcode `\^12\catcode `\_12\catcode `\%12\relax}%
\providecommand \@@startlink[1]{}%
\providecommand \@@endlink[0]{}%
\providecommand \url  [0]{\begingroup\@sanitize@url \@url }%
\providecommand \@url [1]{\endgroup\@href {#1}{\urlprefix }}%
\providecommand \urlprefix  [0]{URL }%
\providecommand \Eprint [0]{\href }%
\providecommand \doibase [0]{http://dx.doi.org/}%
\providecommand \selectlanguage [0]{\@gobble}%
\providecommand \bibinfo  [0]{\@secondoftwo}%
\providecommand \bibfield  [0]{\@secondoftwo}%
\providecommand \translation [1]{[#1]}%
\providecommand \BibitemOpen [0]{}%
\providecommand \bibitemStop [0]{}%
\providecommand \bibitemNoStop [0]{.\EOS\space}%
\providecommand \EOS [0]{\spacefactor3000\relax}%
\providecommand \BibitemShut  [1]{\csname bibitem#1\endcsname}%
\let\auto@bib@innerbib\@empty
\bibitem [{\citenamefont {Sch{\"a}fer}\ \emph {et~al.}(2020)\citenamefont
  {Sch{\"a}fer}, \citenamefont {Fukuhara}, \citenamefont {Sugawa},
  \citenamefont {Takasu},\ and\ \citenamefont {Takahashi}}]{schafer2020}%
  \BibitemOpen
  \bibfield  {author} {\bibinfo {author} {\bibfnamefont {F.}~\bibnamefont
  {Sch{\"a}fer}}, \bibinfo {author} {\bibfnamefont {T.}~\bibnamefont
  {Fukuhara}}, \bibinfo {author} {\bibfnamefont {S.}~\bibnamefont {Sugawa}},
  \bibinfo {author} {\bibfnamefont {Y.}~\bibnamefont {Takasu}}, \ and\ \bibinfo
  {author} {\bibfnamefont {Y.}~\bibnamefont {Takahashi}},\ }\href@noop {}
  {\bibfield  {journal} {\bibinfo  {journal} {Nature Reviews Physics}\ }\textbf
  {\bibinfo {volume} {2}},\ \bibinfo {pages} {411} (\bibinfo {year}
  {2020})}\BibitemShut {NoStop}%
\bibitem [{\citenamefont {Feynman}(1982)}]{feynman}%
  \BibitemOpen
  \bibfield  {author} {\bibinfo {author} {\bibfnamefont {R.~P.}\ \bibnamefont
  {Feynman}},\ }\href@noop {} {\bibfield  {journal} {\bibinfo  {journal} {Int.
  J. Theor. Phys.}\ }\textbf {\bibinfo {volume} {21}},\ \bibinfo {pages} {467}
  (\bibinfo {year} {1982})}\BibitemShut {NoStop}%
\bibitem [{\citenamefont {Bloch}\ \emph {et~al.}(2012)\citenamefont {Bloch},
  \citenamefont {Dalibard},\ and\ \citenamefont {Nascimbene}}]{bloch2012}%
  \BibitemOpen
  \bibfield  {author} {\bibinfo {author} {\bibfnamefont {I.}~\bibnamefont
  {Bloch}}, \bibinfo {author} {\bibfnamefont {J.}~\bibnamefont {Dalibard}}, \
  and\ \bibinfo {author} {\bibfnamefont {S.}~\bibnamefont {Nascimbene}},\
  }\href@noop {} {\bibfield  {journal} {\bibinfo  {journal} {Nature Physics}\
  }\textbf {\bibinfo {volume} {8}},\ \bibinfo {pages} {267} (\bibinfo {year}
  {2012})}\BibitemShut {NoStop}%
\bibitem [{\citenamefont {Blatt}\ and\ \citenamefont {Roos}(2012)}]{blatt2012}%
  \BibitemOpen
  \bibfield  {author} {\bibinfo {author} {\bibfnamefont {R.}~\bibnamefont
  {Blatt}}\ and\ \bibinfo {author} {\bibfnamefont {C.~F.}\ \bibnamefont
  {Roos}},\ }\href@noop {} {\bibfield  {journal} {\bibinfo  {journal} {Nature
  Physics}\ }\textbf {\bibinfo {volume} {8}},\ \bibinfo {pages} {277} (\bibinfo
  {year} {2012})}\BibitemShut {NoStop}%
\bibitem [{\citenamefont {Aspuru-Guzik}\ and\ \citenamefont
  {Walther}(2012)}]{aspuru2012}%
  \BibitemOpen
  \bibfield  {author} {\bibinfo {author} {\bibfnamefont {A.}~\bibnamefont
  {Aspuru-Guzik}}\ and\ \bibinfo {author} {\bibfnamefont {P.}~\bibnamefont
  {Walther}},\ }\href@noop {} {\bibfield  {journal} {\bibinfo  {journal}
  {Nature physics}\ }\textbf {\bibinfo {volume} {8}},\ \bibinfo {pages} {285}
  (\bibinfo {year} {2012})}\BibitemShut {NoStop}%
\bibitem [{\citenamefont {Houck}\ \emph {et~al.}(2012)\citenamefont {Houck},
  \citenamefont {T{\"u}reci},\ and\ \citenamefont {Koch}}]{houck2012}%
  \BibitemOpen
  \bibfield  {author} {\bibinfo {author} {\bibfnamefont {A.~A.}\ \bibnamefont
  {Houck}}, \bibinfo {author} {\bibfnamefont {H.~E.}\ \bibnamefont
  {T{\"u}reci}}, \ and\ \bibinfo {author} {\bibfnamefont {J.}~\bibnamefont
  {Koch}},\ }\href@noop {} {\bibfield  {journal} {\bibinfo  {journal} {Nature
  physics}\ }\textbf {\bibinfo {volume} {8}},\ \bibinfo {pages} {292} (\bibinfo
  {year} {2012})}\BibitemShut {NoStop}%
\bibitem [{\citenamefont {Georgescu}\ \emph {et~al.}(2014)\citenamefont
  {Georgescu}, \citenamefont {Ashhab},\ and\ \citenamefont
  {Nori}}]{RevModPhys.86.153}%
  \BibitemOpen
  \bibfield  {author} {\bibinfo {author} {\bibfnamefont {I.~M.}\ \bibnamefont
  {Georgescu}}, \bibinfo {author} {\bibfnamefont {S.}~\bibnamefont {Ashhab}}, \
  and\ \bibinfo {author} {\bibfnamefont {F.}~\bibnamefont {Nori}},\ }\href
  {\doibase 10.1103/RevModPhys.86.153} {\bibfield  {journal} {\bibinfo
  {journal} {Rev. Mod. Phys.}\ }\textbf {\bibinfo {volume} {86}},\ \bibinfo
  {pages} {153} (\bibinfo {year} {2014})}\BibitemShut {NoStop}%
\bibitem [{\citenamefont {Ueda}(2020)}]{ueda2020}%
  \BibitemOpen
  \bibfield  {author} {\bibinfo {author} {\bibfnamefont {M.}~\bibnamefont
  {Ueda}},\ }\href@noop {} {\bibfield  {journal} {\bibinfo  {journal} {Nature
  Reviews Physics}\ }\textbf {\bibinfo {volume} {2}},\ \bibinfo {pages} {669}
  (\bibinfo {year} {2020})}\BibitemShut {NoStop}%
\bibitem [{\citenamefont {Syassen}\ \emph {et~al.}(2008)\citenamefont
  {Syassen}, \citenamefont {Bauer}, \citenamefont {Lettner}, \citenamefont
  {Volz}, \citenamefont {Dietze}, \citenamefont {Garcia-Ripoll}, \citenamefont
  {Cirac}, \citenamefont {Rempe},\ and\ \citenamefont {Durr}}]{syassen2008}%
  \BibitemOpen
  \bibfield  {author} {\bibinfo {author} {\bibfnamefont {N.}~\bibnamefont
  {Syassen}}, \bibinfo {author} {\bibfnamefont {D.~M.}\ \bibnamefont {Bauer}},
  \bibinfo {author} {\bibfnamefont {M.}~\bibnamefont {Lettner}}, \bibinfo
  {author} {\bibfnamefont {T.}~\bibnamefont {Volz}}, \bibinfo {author}
  {\bibfnamefont {D.}~\bibnamefont {Dietze}}, \bibinfo {author} {\bibfnamefont
  {J.~J.}\ \bibnamefont {Garcia-Ripoll}}, \bibinfo {author} {\bibfnamefont
  {J.~I.}\ \bibnamefont {Cirac}}, \bibinfo {author} {\bibfnamefont
  {G.}~\bibnamefont {Rempe}}, \ and\ \bibinfo {author} {\bibfnamefont
  {S.}~\bibnamefont {Durr}},\ }\href@noop {} {\bibfield  {journal} {\bibinfo
  {journal} {Science}\ }\textbf {\bibinfo {volume} {320}},\ \bibinfo {pages}
  {1329} (\bibinfo {year} {2008})}\BibitemShut {NoStop}%
\bibitem [{\citenamefont {Barontini}\ \emph {et~al.}(2013)\citenamefont
  {Barontini}, \citenamefont {Labouvie}, \citenamefont {Stubenrauch},
  \citenamefont {Vogler}, \citenamefont {Guarrera},\ and\ \citenamefont
  {Ott}}]{PhysRevLett.110.035302}%
  \BibitemOpen
  \bibfield  {author} {\bibinfo {author} {\bibfnamefont {G.}~\bibnamefont
  {Barontini}}, \bibinfo {author} {\bibfnamefont {R.}~\bibnamefont {Labouvie}},
  \bibinfo {author} {\bibfnamefont {F.}~\bibnamefont {Stubenrauch}}, \bibinfo
  {author} {\bibfnamefont {A.}~\bibnamefont {Vogler}}, \bibinfo {author}
  {\bibfnamefont {V.}~\bibnamefont {Guarrera}}, \ and\ \bibinfo {author}
  {\bibfnamefont {H.}~\bibnamefont {Ott}},\ }\href {\doibase
  10.1103/PhysRevLett.110.035302} {\bibfield  {journal} {\bibinfo  {journal}
  {Phys. Rev. Lett.}\ }\textbf {\bibinfo {volume} {110}},\ \bibinfo {pages}
  {035302} (\bibinfo {year} {2013})}\BibitemShut {NoStop}%
\bibitem [{\citenamefont {Patil}\ \emph {et~al.}(2015)\citenamefont {Patil},
  \citenamefont {Chakram},\ and\ \citenamefont
  {Vengalattore}}]{PhysRevLett.115.140402}%
  \BibitemOpen
  \bibfield  {author} {\bibinfo {author} {\bibfnamefont {Y.~S.}\ \bibnamefont
  {Patil}}, \bibinfo {author} {\bibfnamefont {S.}~\bibnamefont {Chakram}}, \
  and\ \bibinfo {author} {\bibfnamefont {M.}~\bibnamefont {Vengalattore}},\
  }\href {\doibase 10.1103/PhysRevLett.115.140402} {\bibfield  {journal}
  {\bibinfo  {journal} {Phys. Rev. Lett.}\ }\textbf {\bibinfo {volume} {115}},\
  \bibinfo {pages} {140402} (\bibinfo {year} {2015})}\BibitemShut {NoStop}%
\bibitem [{\citenamefont {Tomita}\ \emph {et~al.}(2017)\citenamefont {Tomita},
  \citenamefont {Nakajima}, \citenamefont {Danshita}, \citenamefont {Takasu},\
  and\ \citenamefont {Takahashi}}]{tomita2017}%
  \BibitemOpen
  \bibfield  {author} {\bibinfo {author} {\bibfnamefont {T.}~\bibnamefont
  {Tomita}}, \bibinfo {author} {\bibfnamefont {S.}~\bibnamefont {Nakajima}},
  \bibinfo {author} {\bibfnamefont {I.}~\bibnamefont {Danshita}}, \bibinfo
  {author} {\bibfnamefont {Y.}~\bibnamefont {Takasu}}, \ and\ \bibinfo {author}
  {\bibfnamefont {Y.}~\bibnamefont {Takahashi}},\ }\href@noop {} {\bibfield
  {journal} {\bibinfo  {journal} {Science advances}\ }\textbf {\bibinfo
  {volume} {3}},\ \bibinfo {pages} {e1701513} (\bibinfo {year}
  {2017})}\BibitemShut {NoStop}%
\bibitem [{\citenamefont {Schemmer}\ and\ \citenamefont
  {Bouchoule}(2018)}]{PhysRevLett.121.200401}%
  \BibitemOpen
  \bibfield  {author} {\bibinfo {author} {\bibfnamefont {M.}~\bibnamefont
  {Schemmer}}\ and\ \bibinfo {author} {\bibfnamefont {I.}~\bibnamefont
  {Bouchoule}},\ }\href {\doibase 10.1103/PhysRevLett.121.200401} {\bibfield
  {journal} {\bibinfo  {journal} {Phys. Rev. Lett.}\ }\textbf {\bibinfo
  {volume} {121}},\ \bibinfo {pages} {200401} (\bibinfo {year}
  {2018})}\BibitemShut {NoStop}%
\bibitem [{\citenamefont {Corman}\ \emph {et~al.}(2019)\citenamefont {Corman},
  \citenamefont {Fabritius}, \citenamefont {H\"ausler}, \citenamefont {Mohan},
  \citenamefont {Dogra}, \citenamefont {Husmann}, \citenamefont {Lebrat},\ and\
  \citenamefont {Esslinger}}]{PhysRevA.100.053605}%
  \BibitemOpen
  \bibfield  {author} {\bibinfo {author} {\bibfnamefont {L.}~\bibnamefont
  {Corman}}, \bibinfo {author} {\bibfnamefont {P.}~\bibnamefont {Fabritius}},
  \bibinfo {author} {\bibfnamefont {S.}~\bibnamefont {H\"ausler}}, \bibinfo
  {author} {\bibfnamefont {J.}~\bibnamefont {Mohan}}, \bibinfo {author}
  {\bibfnamefont {L.~H.}\ \bibnamefont {Dogra}}, \bibinfo {author}
  {\bibfnamefont {D.}~\bibnamefont {Husmann}}, \bibinfo {author} {\bibfnamefont
  {M.}~\bibnamefont {Lebrat}}, \ and\ \bibinfo {author} {\bibfnamefont
  {T.}~\bibnamefont {Esslinger}},\ }\href {\doibase
  10.1103/PhysRevA.100.053605} {\bibfield  {journal} {\bibinfo  {journal}
  {Phys. Rev. A}\ }\textbf {\bibinfo {volume} {100}},\ \bibinfo {pages}
  {053605} (\bibinfo {year} {2019})}\BibitemShut {NoStop}%
\bibitem [{\citenamefont {Konishi}\ \emph {et~al.}(2021)\citenamefont
  {Konishi}, \citenamefont {Roux}, \citenamefont {Helson},\ and\ \citenamefont
  {Brantut}}]{konishi2021}%
  \BibitemOpen
  \bibfield  {author} {\bibinfo {author} {\bibfnamefont {H.}~\bibnamefont
  {Konishi}}, \bibinfo {author} {\bibfnamefont {K.}~\bibnamefont {Roux}},
  \bibinfo {author} {\bibfnamefont {V.}~\bibnamefont {Helson}}, \ and\ \bibinfo
  {author} {\bibfnamefont {J.-P.}\ \bibnamefont {Brantut}},\ }\href@noop {}
  {\bibfield  {journal} {\bibinfo  {journal} {Nature}\ }\textbf {\bibinfo
  {volume} {596}},\ \bibinfo {pages} {509} (\bibinfo {year}
  {2021})}\BibitemShut {NoStop}%
\bibitem [{\citenamefont {Datta}(1997)}]{datta1997}%
  \BibitemOpen
  \bibfield  {author} {\bibinfo {author} {\bibfnamefont {S.}~\bibnamefont
  {Datta}},\ }\href@noop {} {\emph {\bibinfo {title} {Electronic transport in
  mesoscopic systems}}}\ (\bibinfo  {publisher} {Cambridge university press},\
  \bibinfo {year} {1997})\BibitemShut {NoStop}%
\bibitem [{\citenamefont {Nazarov}\ and\ \citenamefont
  {Blanter}(2009)}]{nazarov}%
  \BibitemOpen
  \bibfield  {author} {\bibinfo {author} {\bibfnamefont {Y.~V.}\ \bibnamefont
  {Nazarov}}\ and\ \bibinfo {author} {\bibfnamefont {Y.~M.}\ \bibnamefont
  {Blanter}},\ }\href@noop {} {\emph {\bibinfo {title} {Quantum transport:
  introduction to nanoscience}}}\ (\bibinfo  {publisher} {Cambridge University
  Press},\ \bibinfo {year} {2009})\BibitemShut {NoStop}%
\bibitem [{\citenamefont {Labouvie}\ \emph {et~al.}(2015)\citenamefont
  {Labouvie}, \citenamefont {Santra}, \citenamefont {Heun}, \citenamefont
  {Wimberger},\ and\ \citenamefont {Ott}}]{PhysRevLett.115.050601}%
  \BibitemOpen
  \bibfield  {author} {\bibinfo {author} {\bibfnamefont {R.}~\bibnamefont
  {Labouvie}}, \bibinfo {author} {\bibfnamefont {B.}~\bibnamefont {Santra}},
  \bibinfo {author} {\bibfnamefont {S.}~\bibnamefont {Heun}}, \bibinfo {author}
  {\bibfnamefont {S.}~\bibnamefont {Wimberger}}, \ and\ \bibinfo {author}
  {\bibfnamefont {H.}~\bibnamefont {Ott}},\ }\href {\doibase
  10.1103/PhysRevLett.115.050601} {\bibfield  {journal} {\bibinfo  {journal}
  {Phys. Rev. Lett.}\ }\textbf {\bibinfo {volume} {115}},\ \bibinfo {pages}
  {050601} (\bibinfo {year} {2015})}\BibitemShut {NoStop}%
\bibitem [{\citenamefont {Labouvie}\ \emph {et~al.}(2016)\citenamefont
  {Labouvie}, \citenamefont {Santra}, \citenamefont {Heun},\ and\ \citenamefont
  {Ott}}]{PhysRevLett.116.235302}%
  \BibitemOpen
  \bibfield  {author} {\bibinfo {author} {\bibfnamefont {R.}~\bibnamefont
  {Labouvie}}, \bibinfo {author} {\bibfnamefont {B.}~\bibnamefont {Santra}},
  \bibinfo {author} {\bibfnamefont {S.}~\bibnamefont {Heun}}, \ and\ \bibinfo
  {author} {\bibfnamefont {H.}~\bibnamefont {Ott}},\ }\href {\doibase
  10.1103/PhysRevLett.116.235302} {\bibfield  {journal} {\bibinfo  {journal}
  {Phys. Rev. Lett.}\ }\textbf {\bibinfo {volume} {116}},\ \bibinfo {pages}
  {235302} (\bibinfo {year} {2016})}\BibitemShut {NoStop}%
\bibitem [{\citenamefont {Benary}\ \emph {et~al.}(2022)\citenamefont {Benary},
  \citenamefont {Baals}, \citenamefont {Bernhart}, \citenamefont {Jiang},
  \citenamefont {R{\"o}hrle},\ and\ \citenamefont {Ott}}]{benary2022}%
  \BibitemOpen
  \bibfield  {author} {\bibinfo {author} {\bibfnamefont {J.}~\bibnamefont
  {Benary}}, \bibinfo {author} {\bibfnamefont {C.}~\bibnamefont {Baals}},
  \bibinfo {author} {\bibfnamefont {E.}~\bibnamefont {Bernhart}}, \bibinfo
  {author} {\bibfnamefont {J.}~\bibnamefont {Jiang}}, \bibinfo {author}
  {\bibfnamefont {M.}~\bibnamefont {R{\"o}hrle}}, \ and\ \bibinfo {author}
  {\bibfnamefont {H.}~\bibnamefont {Ott}},\ }\href@noop {} {\bibfield
  {journal} {\bibinfo  {journal} {arXiv preprint arXiv:2203.09896}\ } (\bibinfo
  {year} {2022})}\BibitemShut {NoStop}%
\bibitem [{\citenamefont {Daley}(2014)}]{daley2014}%
  \BibitemOpen
  \bibfield  {author} {\bibinfo {author} {\bibfnamefont {A.~J.}\ \bibnamefont
  {Daley}},\ }\href@noop {} {\bibfield  {journal} {\bibinfo  {journal}
  {Advances in Physics}\ }\textbf {\bibinfo {volume} {63}},\ \bibinfo {pages}
  {77} (\bibinfo {year} {2014})}\BibitemShut {NoStop}%
\bibitem [{\citenamefont {Moiseyev}(2011)}]{moiseyev2011}%
  \BibitemOpen
  \bibfield  {author} {\bibinfo {author} {\bibfnamefont {N.}~\bibnamefont
  {Moiseyev}},\ }\href@noop {} {\emph {\bibinfo {title} {Non-Hermitian quantum
  mechanics}}}\ (\bibinfo  {publisher} {Cambridge University Press},\ \bibinfo
  {year} {2011})\BibitemShut {NoStop}%
\bibitem [{\citenamefont {Ashida}\ \emph {et~al.}(2020)\citenamefont {Ashida},
  \citenamefont {Gong},\ and\ \citenamefont {Ueda}}]{ashida2020}%
  \BibitemOpen
  \bibfield  {author} {\bibinfo {author} {\bibfnamefont {Y.}~\bibnamefont
  {Ashida}}, \bibinfo {author} {\bibfnamefont {Z.}~\bibnamefont {Gong}}, \ and\
  \bibinfo {author} {\bibfnamefont {M.}~\bibnamefont {Ueda}},\ }\href@noop {}
  {\bibfield  {journal} {\bibinfo  {journal} {Advances in Physics}\ }\textbf
  {\bibinfo {volume} {69}},\ \bibinfo {pages} {249} (\bibinfo {year}
  {2020})}\BibitemShut {NoStop}%
\bibitem [{\citenamefont {Visuri}\ \emph {et~al.}(2022)\citenamefont {Visuri},
  \citenamefont {Giamarchi},\ and\ \citenamefont {Kollath}}]{visuri2022}%
  \BibitemOpen
  \bibfield  {author} {\bibinfo {author} {\bibfnamefont {A.-M.}\ \bibnamefont
  {Visuri}}, \bibinfo {author} {\bibfnamefont {T.}~\bibnamefont {Giamarchi}}, \
  and\ \bibinfo {author} {\bibfnamefont {C.}~\bibnamefont {Kollath}},\ }\href
  {\doibase 10.1103/PhysRevLett.129.056802} {\bibfield  {journal} {\bibinfo
  {journal} {Phys. Rev. Lett.}\ }\textbf {\bibinfo {volume} {129}},\ \bibinfo
  {pages} {056802} (\bibinfo {year} {2022})}\BibitemShut {NoStop}%
\bibitem [{\citenamefont {Jin}\ \emph {et~al.}(2020)\citenamefont {Jin},
  \citenamefont {Filippone},\ and\ \citenamefont
  {Giamarchi}}]{PhysRevB.102.205131}%
  \BibitemOpen
  \bibfield  {author} {\bibinfo {author} {\bibfnamefont {T.}~\bibnamefont
  {Jin}}, \bibinfo {author} {\bibfnamefont {M.}~\bibnamefont {Filippone}}, \
  and\ \bibinfo {author} {\bibfnamefont {T.}~\bibnamefont {Giamarchi}},\ }\href
  {\doibase 10.1103/PhysRevB.102.205131} {\bibfield  {journal} {\bibinfo
  {journal} {Phys. Rev. B}\ }\textbf {\bibinfo {volume} {102}},\ \bibinfo
  {pages} {205131} (\bibinfo {year} {2020})}\BibitemShut {NoStop}%
\bibitem [{\citenamefont {Sieberer}\ \emph {et~al.}(2016)\citenamefont
  {Sieberer}, \citenamefont {Buchhold},\ and\ \citenamefont
  {Diehl}}]{sieberer2016}%
  \BibitemOpen
  \bibfield  {author} {\bibinfo {author} {\bibfnamefont {L.~M.}\ \bibnamefont
  {Sieberer}}, \bibinfo {author} {\bibfnamefont {M.}~\bibnamefont {Buchhold}},
  \ and\ \bibinfo {author} {\bibfnamefont {S.}~\bibnamefont {Diehl}},\
  }\href@noop {} {\bibfield  {journal} {\bibinfo  {journal} {Reports on
  Progress in Physics}\ }\textbf {\bibinfo {volume} {79}},\ \bibinfo {pages}
  {096001} (\bibinfo {year} {2016})}\BibitemShut {NoStop}%
\bibitem [{\citenamefont {Breuer}\ \emph {et~al.}(2002)\citenamefont {Breuer},
  \citenamefont {Petruccione} \emph {et~al.}}]{breuer2002}%
  \BibitemOpen
  \bibfield  {author} {\bibinfo {author} {\bibfnamefont {H.-P.}\ \bibnamefont
  {Breuer}}, \bibinfo {author} {\bibfnamefont {F.}~\bibnamefont {Petruccione}},
  \ }\href@noop {} {\emph {\bibinfo {title} {The theory of open
  quantum systems}}}\ (\bibinfo  {publisher} {Oxford University Press on
  Demand},\ \bibinfo {year} {2002})\BibitemShut {NoStop}%
\bibitem [{Note1()}]{Note1}%
  \BibitemOpen
  \bibinfo {note} {In a precise sense, ultracold atomic gases are finite
  systems and therefore the chemical potential in each reservoir is not exactly
  a constant in time. Thus, the constant $\mu _{L(R)}$ treatment becomes a good
  approximation if thermalization in each reservoir is enough fast compared to
  the transport (relaxation) time scale~\cite {krinner2017}.}\BibitemShut
  {Stop}%
\bibitem [{\citenamefont {Kamenev}(2011)}]{kamenev2011}%
  \BibitemOpen
  \bibfield  {author} {\bibinfo {author} {\bibfnamefont {A.}~\bibnamefont
  {Kamenev}},\ }\href@noop {} {\emph {\bibinfo {title} {Field theory of
  non-equilibrium systems}}}\ (\bibinfo  {publisher} {Cambridge University
  Press},\ \bibinfo {year} {2011})\BibitemShut {NoStop}%
\bibitem [{\citenamefont {Bolech}\ and\ \citenamefont
  {Giamarchi}(2004)}]{PhysRevLett.92.127001}%
  \BibitemOpen
  \bibfield  {author} {\bibinfo {author} {\bibfnamefont {C.~J.}\ \bibnamefont
  {Bolech}}\ and\ \bibinfo {author} {\bibfnamefont {T.}~\bibnamefont
  {Giamarchi}},\ }\href {\doibase 10.1103/PhysRevLett.92.127001} {\bibfield
  {journal} {\bibinfo  {journal} {Phys. Rev. Lett.}\ }\textbf {\bibinfo
  {volume} {92}},\ \bibinfo {pages} {127001} (\bibinfo {year}
  {2004})}\BibitemShut {NoStop}%
\bibitem [{\citenamefont {Bolech}\ and\ \citenamefont
  {Giamarchi}(2005)}]{PhysRevB.71.024517}%
  \BibitemOpen
  \bibfield  {author} {\bibinfo {author} {\bibfnamefont {C.~J.}\ \bibnamefont
  {Bolech}}\ and\ \bibinfo {author} {\bibfnamefont {T.}~\bibnamefont
  {Giamarchi}},\ }\href {\doibase 10.1103/PhysRevB.71.024517} {\bibfield
  {journal} {\bibinfo  {journal} {Phys. Rev. B}\ }\textbf {\bibinfo {volume}
  {71}},\ \bibinfo {pages} {024517} (\bibinfo {year} {2005})}\BibitemShut
  {NoStop}%
\bibitem [{\citenamefont {Husmann}\ \emph {et~al.}(2015)\citenamefont
  {Husmann}, \citenamefont {Uchino}, \citenamefont {Krinner}, \citenamefont
  {Lebrat}, \citenamefont {Giamarchi}, \citenamefont {Esslinger},\ and\
  \citenamefont {Brantut}}]{husmann2015}%
  \BibitemOpen
  \bibfield  {author} {\bibinfo {author} {\bibfnamefont {D.}~\bibnamefont
  {Husmann}}, \bibinfo {author} {\bibfnamefont {S.}~\bibnamefont {Uchino}},
  \bibinfo {author} {\bibfnamefont {S.}~\bibnamefont {Krinner}}, \bibinfo
  {author} {\bibfnamefont {M.}~\bibnamefont {Lebrat}}, \bibinfo {author}
  {\bibfnamefont {T.}~\bibnamefont {Giamarchi}}, \bibinfo {author}
  {\bibfnamefont {T.}~\bibnamefont {Esslinger}}, \ and\ \bibinfo {author}
  {\bibfnamefont {J.-P.}\ \bibnamefont {Brantut}},\ }\href@noop {} {\bibfield
  {journal} {\bibinfo  {journal} {Science}\ }\textbf {\bibinfo {volume}
  {350}},\ \bibinfo {pages} {1498} (\bibinfo {year} {2015})}\BibitemShut
  {NoStop}%
\bibitem [{\citenamefont {Yao}\ \emph {et~al.}(2018)\citenamefont {Yao},
  \citenamefont {Liu}, \citenamefont {Sun},\ and\ \citenamefont
  {Zhai}}]{PhysRevA.98.041601}%
  \BibitemOpen
  \bibfield  {author} {\bibinfo {author} {\bibfnamefont {J.}~\bibnamefont
  {Yao}}, \bibinfo {author} {\bibfnamefont {B.}~\bibnamefont {Liu}}, \bibinfo
  {author} {\bibfnamefont {M.}~\bibnamefont {Sun}}, \ and\ \bibinfo {author}
  {\bibfnamefont {H.}~\bibnamefont {Zhai}},\ }\href {\doibase
  10.1103/PhysRevA.98.041601} {\bibfield  {journal} {\bibinfo  {journal} {Phys.
  Rev. A}\ }\textbf {\bibinfo {volume} {98}},\ \bibinfo {pages} {041601}
  (\bibinfo {year} {2018})}\BibitemShut {NoStop}%
\bibitem [{\citenamefont {Rammer}(2007)}]{rammer2007}%
  \BibitemOpen
  \bibfield  {author} {\bibinfo {author} {\bibfnamefont {J.}~\bibnamefont
  {Rammer}},\ }\href@noop {} {\emph {\bibinfo {title} {Quantum field theory of
  non-equilibrium states}}}\ (\bibinfo  {publisher} {Cambridge University
  Press},\ \bibinfo {year} {2007})\BibitemShut {NoStop}%
\bibitem [{\citenamefont {Haug}\ \emph {et~al.}(2008)\citenamefont {Haug},
  \citenamefont {Jauho} \emph {et~al.}}]{haug2008}%
  \BibitemOpen
  \bibfield  {author} {\bibinfo {author} {\bibfnamefont {H.}~\bibnamefont
  {Haug}}, \bibinfo {author} {\bibfnamefont {A.-P.}\ \bibnamefont {Jauho}},
   }\href@noop {} {\emph {\bibinfo {title} {Quantum kinetics in
  transport and optics of semiconductors}}},\ Vol.~\bibinfo {volume} {2}\
  (\bibinfo  {publisher} {Springer},\ \bibinfo {year} {2008})\BibitemShut
  {NoStop}%
\bibitem [{\citenamefont {Meir}\ and\ \citenamefont
  {Wingreen}(1992)}]{PhysRevLett.68.2512}%
  \BibitemOpen
  \bibfield  {author} {\bibinfo {author} {\bibfnamefont {Y.}~\bibnamefont
  {Meir}}\ and\ \bibinfo {author} {\bibfnamefont {N.~S.}\ \bibnamefont
  {Wingreen}},\ }\href {\doibase 10.1103/PhysRevLett.68.2512} {\bibfield
  {journal} {\bibinfo  {journal} {Phys. Rev. Lett.}\ }\textbf {\bibinfo
  {volume} {68}},\ \bibinfo {pages} {2512} (\bibinfo {year}
  {1992})}\BibitemShut {NoStop}%
\bibitem [{\citenamefont {Jonson}\ and\ \citenamefont
  {Grincwajg}(1987)}]{jonson1987}%
  \BibitemOpen
  \bibfield  {author} {\bibinfo {author} {\bibfnamefont {M.}~\bibnamefont
  {Jonson}}\ and\ \bibinfo {author} {\bibfnamefont {A.}~\bibnamefont
  {Grincwajg}},\ }\href@noop {} {\bibfield  {journal} {\bibinfo  {journal}
  {Applied physics letters}\ }\textbf {\bibinfo {volume} {51}},\ \bibinfo
  {pages} {1729} (\bibinfo {year} {1987})}\BibitemShut {NoStop}%
\bibitem [{\citenamefont {Berthod}\ and\ \citenamefont
  {Giamarchi}(2011)}]{PhysRevB.84.155414}%
  \BibitemOpen
  \bibfield  {author} {\bibinfo {author} {\bibfnamefont {C.}~\bibnamefont
  {Berthod}}\ and\ \bibinfo {author} {\bibfnamefont {T.}~\bibnamefont
  {Giamarchi}},\ }\href {\doibase 10.1103/PhysRevB.84.155414} {\bibfield
  {journal} {\bibinfo  {journal} {Phys. Rev. B}\ }\textbf {\bibinfo {volume}
  {84}},\ \bibinfo {pages} {155414} (\bibinfo {year} {2011})}\BibitemShut
  {NoStop}%
\bibitem [{\citenamefont {Uchino}\ and\ \citenamefont
  {Ueda}(2017)}]{PhysRevLett.118.105303}%
  \BibitemOpen
  \bibfield  {author} {\bibinfo {author} {\bibfnamefont {S.}~\bibnamefont
  {Uchino}}\ and\ \bibinfo {author} {\bibfnamefont {M.}~\bibnamefont {Ueda}},\
  }\href {\doibase 10.1103/PhysRevLett.118.105303} {\bibfield  {journal}
  {\bibinfo  {journal} {Phys. Rev. Lett.}\ }\textbf {\bibinfo {volume} {118}},\
  \bibinfo {pages} {105303} (\bibinfo {year} {2017})}\BibitemShut {NoStop}%
\bibitem [{\citenamefont {Uchino}\ and\ \citenamefont
  {Brantut}(2020)}]{PhysRevResearch.2.023284}%
  \BibitemOpen
  \bibfield  {author} {\bibinfo {author} {\bibfnamefont {S.}~\bibnamefont
  {Uchino}}\ and\ \bibinfo {author} {\bibfnamefont {J.-P.}\ \bibnamefont
  {Brantut}},\ }\href {\doibase 10.1103/PhysRevResearch.2.023284} {\bibfield
  {journal} {\bibinfo  {journal} {Phys. Rev. Research}\ }\textbf {\bibinfo
  {volume} {2}},\ \bibinfo {pages} {023284} (\bibinfo {year}
  {2020})}\BibitemShut {NoStop}%
\bibitem [{\citenamefont {Uchino}(2021)}]{PhysRevResearch.3.043058}%
  \BibitemOpen
  \bibfield  {author} {\bibinfo {author} {\bibfnamefont {S.}~\bibnamefont
  {Uchino}},\ }\href {\doibase 10.1103/PhysRevResearch.3.043058} {\bibfield
  {journal} {\bibinfo  {journal} {Phys. Rev. Research}\ }\textbf {\bibinfo
  {volume} {3}},\ \bibinfo {pages} {043058} (\bibinfo {year}
  {2021})}\BibitemShut {NoStop}%
\bibitem [{\citenamefont {Uchino}(2022)}]{PhysRevA.106.L011303}%
  \BibitemOpen
  \bibfield  {author} {\bibinfo {author} {\bibfnamefont {S.}~\bibnamefont
  {Uchino}},\ }\href {\doibase 10.1103/PhysRevA.106.L011303} {\bibfield
  {journal} {\bibinfo  {journal} {Phys. Rev. A}\ }\textbf {\bibinfo {volume}
  {106}},\ \bibinfo {pages} {L011303} (\bibinfo {year} {2022})}\BibitemShut
  {NoStop}%
\bibitem [{\citenamefont {Da~Fonseca}(2007)}]{da2007}%
  \BibitemOpen
  \bibfield  {author} {\bibinfo {author} {\bibfnamefont {C.}~\bibnamefont
  {Da~Fonseca}},\ }\href@noop {} {\bibfield  {journal} {\bibinfo  {journal}
  {Journal of Computational and Applied Mathematics}\ }\textbf {\bibinfo
  {volume} {200}},\ \bibinfo {pages} {283} (\bibinfo {year}
  {2007})}\BibitemShut {NoStop}%
\bibitem [{\citenamefont {Usmani}(1994)}]{usmani1994}%
  \BibitemOpen
  \bibfield  {author} {\bibinfo {author} {\bibfnamefont {R.~A.}\ \bibnamefont
  {Usmani}},\ }\href@noop {} {\bibfield  {journal} {\bibinfo  {journal} {Linear
  Algebra and its Applications}\ }\textbf {\bibinfo {volume} {212}},\ \bibinfo
  {pages} {413} (\bibinfo {year} {1994})}\BibitemShut {NoStop}%
\bibitem [{\citenamefont {Turkeshi}\ and\ \citenamefont
  {Schir\'o}(2021)}]{PhysRevB.104.144301}%
  \BibitemOpen
  \bibfield  {author} {\bibinfo {author} {\bibfnamefont {X.}~\bibnamefont
  {Turkeshi}}\ and\ \bibinfo {author} {\bibfnamefont {M.}~\bibnamefont
  {Schir\'o}},\ }\href {\doibase 10.1103/PhysRevB.104.144301} {\bibfield
  {journal} {\bibinfo  {journal} {Phys. Rev. B}\ }\textbf {\bibinfo {volume}
  {104}},\ \bibinfo {pages} {144301} (\bibinfo {year} {2021})}\BibitemShut
  {NoStop}%
\bibitem [{\citenamefont {Jin}\ \emph {et~al.}(2022)\citenamefont {Jin},
  \citenamefont {Ferreira}, \citenamefont {Filippone},\ and\ \citenamefont
  {Giamarchi}}]{PhysRevResearch.4.013109}%
  \BibitemOpen
  \bibfield  {author} {\bibinfo {author} {\bibfnamefont {T.}~\bibnamefont
  {Jin}}, \bibinfo {author} {\bibfnamefont {J.~a.~S.}\ \bibnamefont
  {Ferreira}}, \bibinfo {author} {\bibfnamefont {M.}~\bibnamefont {Filippone}},
  \ and\ \bibinfo {author} {\bibfnamefont {T.}~\bibnamefont {Giamarchi}},\
  }\href {\doibase 10.1103/PhysRevResearch.4.013109} {\bibfield  {journal}
  {\bibinfo  {journal} {Phys. Rev. Research}\ }\textbf {\bibinfo {volume}
  {4}},\ \bibinfo {pages} {013109} (\bibinfo {year} {2022})}\BibitemShut
  {NoStop}%
\bibitem [{\citenamefont {B\"uttiker}\ \emph {et~al.}(1984)\citenamefont
  {B\"uttiker}, \citenamefont {Imry},\ and\ \citenamefont
  {Azbel}}]{PhysRevA.30.1982}%
  \BibitemOpen
  \bibfield  {author} {\bibinfo {author} {\bibfnamefont {M.}~\bibnamefont
  {B\"uttiker}}, \bibinfo {author} {\bibfnamefont {Y.}~\bibnamefont {Imry}}, \
  and\ \bibinfo {author} {\bibfnamefont {M.~Y.}\ \bibnamefont {Azbel}},\ }\href
  {\doibase 10.1103/PhysRevA.30.1982} {\bibfield  {journal} {\bibinfo
  {journal} {Phys. Rev. A}\ }\textbf {\bibinfo {volume} {30}},\ \bibinfo
  {pages} {1982} (\bibinfo {year} {1984})}\BibitemShut {NoStop}%
\bibitem [{\citenamefont {Fr\"oml}\ \emph {et~al.}(2019)\citenamefont
  {Fr\"oml}, \citenamefont {Chiocchetta}, \citenamefont {Kollath},\ and\
  \citenamefont {Diehl}}]{PhysRevLett.122.040402}%
  \BibitemOpen
  \bibfield  {author} {\bibinfo {author} {\bibfnamefont {H.}~\bibnamefont
  {Fr\"oml}}, \bibinfo {author} {\bibfnamefont {A.}~\bibnamefont
  {Chiocchetta}}, \bibinfo {author} {\bibfnamefont {C.}~\bibnamefont
  {Kollath}}, \ and\ \bibinfo {author} {\bibfnamefont {S.}~\bibnamefont
  {Diehl}},\ }\href {\doibase 10.1103/PhysRevLett.122.040402} {\bibfield
  {journal} {\bibinfo  {journal} {Phys. Rev. Lett.}\ }\textbf {\bibinfo
  {volume} {122}},\ \bibinfo {pages} {040402} (\bibinfo {year}
  {2019})}\BibitemShut {NoStop}%
\bibitem [{\citenamefont {Fr\"oml}\ \emph {et~al.}(2020)\citenamefont
  {Fr\"oml}, \citenamefont {Muckel}, \citenamefont {Kollath}, \citenamefont
  {Chiocchetta},\ and\ \citenamefont {Diehl}}]{PhysRevB.101.144301}%
  \BibitemOpen
  \bibfield  {author} {\bibinfo {author} {\bibfnamefont {H.}~\bibnamefont
  {Fr\"oml}}, \bibinfo {author} {\bibfnamefont {C.}~\bibnamefont {Muckel}},
  \bibinfo {author} {\bibfnamefont {C.}~\bibnamefont {Kollath}}, \bibinfo
  {author} {\bibfnamefont {A.}~\bibnamefont {Chiocchetta}}, \ and\ \bibinfo
  {author} {\bibfnamefont {S.}~\bibnamefont {Diehl}},\ }\href {\doibase
  10.1103/PhysRevB.101.144301} {\bibfield  {journal} {\bibinfo  {journal}
  {Phys. Rev. B}\ }\textbf {\bibinfo {volume} {101}},\ \bibinfo {pages}
  {144301} (\bibinfo {year} {2020})}\BibitemShut {NoStop}%
\bibitem [{Note2()}]{Note2}%
  \BibitemOpen
  \bibinfo {note} {As pointed out in Sec.~III~A, $\omega $ used in this paper
  is measured from the average chemical potential between left and right
  reservoirs. In order to obtain the typical forms adopted in the
  Landauer-Buttiker analyses where $\omega $ corresponds to the absolute
  energy, the energy shift between different $\omega $'s must be taken into
  account.}\BibitemShut {Stop}%
\bibitem [{\citenamefont {B\"uttiker}(1986)}]{PhysRevB.33.3020}%
  \BibitemOpen
  \bibfield  {author} {\bibinfo {author} {\bibfnamefont {M.}~\bibnamefont
  {B\"uttiker}},\ }\href {\doibase 10.1103/PhysRevB.33.3020} {\bibfield
  {journal} {\bibinfo  {journal} {Phys. Rev. B}\ }\textbf {\bibinfo {volume}
  {33}},\ \bibinfo {pages} {3020} (\bibinfo {year} {1986})}\BibitemShut
  {NoStop}%
\bibitem [{\citenamefont {Uchino}(2020)}]{PhysRevResearch.2.023340}%
  \BibitemOpen
  \bibfield  {author} {\bibinfo {author} {\bibfnamefont {S.}~\bibnamefont
  {Uchino}},\ }\href {\doibase 10.1103/PhysRevResearch.2.023340} {\bibfield
  {journal} {\bibinfo  {journal} {Phys. Rev. Research}\ }\textbf {\bibinfo
  {volume} {2}},\ \bibinfo {pages} {023340} (\bibinfo {year}
  {2020})}\BibitemShut {NoStop}%
\bibitem [{\citenamefont {Krinner}\ \emph {et~al.}(2016)\citenamefont
  {Krinner}, \citenamefont {Lebrat}, \citenamefont {Husmann}, \citenamefont
  {Grenier}, \citenamefont {Brantut},\ and\ \citenamefont
  {Esslinger}}]{krinner2016}%
  \BibitemOpen
  \bibfield  {author} {\bibinfo {author} {\bibfnamefont {S.}~\bibnamefont
  {Krinner}}, \bibinfo {author} {\bibfnamefont {M.}~\bibnamefont {Lebrat}},
  \bibinfo {author} {\bibfnamefont {D.}~\bibnamefont {Husmann}}, \bibinfo
  {author} {\bibfnamefont {C.}~\bibnamefont {Grenier}}, \bibinfo {author}
  {\bibfnamefont {J.-P.}\ \bibnamefont {Brantut}}, \ and\ \bibinfo {author}
  {\bibfnamefont {T.}~\bibnamefont {Esslinger}},\ }\href@noop {} {\bibfield
  {journal} {\bibinfo  {journal} {Proceedings of the National Academy of
  Sciences}\ }\textbf {\bibinfo {volume} {113}},\ \bibinfo {pages} {8144}
  (\bibinfo {year} {2016})}\BibitemShut {NoStop}%
\bibitem [{\citenamefont {Kanasz-Nagy}\ \emph {et~al.}(2016)\citenamefont
  {Kanasz-Nagy}, \citenamefont {Glazman}, \citenamefont {Esslinger},\ and\
  \citenamefont {Demler}}]{PhysRevLett.117.255302}%
  \BibitemOpen
  \bibfield  {author} {\bibinfo {author} {\bibfnamefont {M.}~\bibnamefont
  {Kanasz-Nagy}}, \bibinfo {author} {\bibfnamefont {L.}~\bibnamefont
  {Glazman}}, \bibinfo {author} {\bibfnamefont {T.}~\bibnamefont {Esslinger}},
  \ and\ \bibinfo {author} {\bibfnamefont {E.~A.}\ \bibnamefont {Demler}},\
  }\href {\doibase 10.1103/PhysRevLett.117.255302} {\bibfield  {journal}
  {\bibinfo  {journal} {Phys. Rev. Lett.}\ }\textbf {\bibinfo {volume} {117}},\
  \bibinfo {pages} {255302} (\bibinfo {year} {2016})}\BibitemShut {NoStop}%
\bibitem [{\citenamefont {Lebrat}\ \emph {et~al.}(2019)\citenamefont {Lebrat},
  \citenamefont {Hausler}, \citenamefont {Fabritius}, \citenamefont {Husmann},
  \citenamefont {Corman},\ and\ \citenamefont
  {Esslinger}}]{PhysRevLett.123.193605}%
  \BibitemOpen
  \bibfield  {author} {\bibinfo {author} {\bibfnamefont {M.}~\bibnamefont
  {Lebrat}}, \bibinfo {author} {\bibfnamefont {S.}~\bibnamefont {Hausler}},
  \bibinfo {author} {\bibfnamefont {P.}~\bibnamefont {Fabritius}}, \bibinfo
  {author} {\bibfnamefont {D.}~\bibnamefont {Husmann}}, \bibinfo {author}
  {\bibfnamefont {L.}~\bibnamefont {Corman}}, \ and\ \bibinfo {author}
  {\bibfnamefont {T.}~\bibnamefont {Esslinger}},\ }\href {\doibase
  10.1103/PhysRevLett.123.193605} {\bibfield  {journal} {\bibinfo  {journal}
  {Phys. Rev. Lett.}\ }\textbf {\bibinfo {volume} {123}},\ \bibinfo {pages}
  {193605} (\bibinfo {year} {2019})}\BibitemShut {NoStop}%
\bibitem [{\citenamefont {Zhang}\ and\ \citenamefont
  {Sommer}(2019)}]{zhang2019}%
  \BibitemOpen
  \bibfield  {author} {\bibinfo {author} {\bibfnamefont {D.}~\bibnamefont
  {Zhang}}\ and\ \bibinfo {author} {\bibfnamefont {A.~T.}\ \bibnamefont
  {Sommer}},\ }\href@noop {} {\bibfield  {journal} {\bibinfo  {journal} {arXiv
  preprint arXiv:1912.06131}\ } (\bibinfo {year} {2019})}\BibitemShut {NoStop}%
\bibitem [{\citenamefont {Damanet}\ \emph
  {et~al.}(2019{\natexlab{a}})\citenamefont {Damanet}, \citenamefont
  {Mascarenhas}, \citenamefont {Pekker},\ and\ \citenamefont
  {Daley}}]{PhysRevLett.123.180402}%
  \BibitemOpen
  \bibfield  {author} {\bibinfo {author} {\bibfnamefont {F.}~\bibnamefont
  {Damanet}}, \bibinfo {author} {\bibfnamefont {E.}~\bibnamefont
  {Mascarenhas}}, \bibinfo {author} {\bibfnamefont {D.}~\bibnamefont {Pekker}},
  \ and\ \bibinfo {author} {\bibfnamefont {A.~J.}\ \bibnamefont {Daley}},\
  }\href {\doibase 10.1103/PhysRevLett.123.180402} {\bibfield  {journal}
  {\bibinfo  {journal} {Phys. Rev. Lett.}\ }\textbf {\bibinfo {volume} {123}},\
  \bibinfo {pages} {180402} (\bibinfo {year} {2019}{\natexlab{a}})}\BibitemShut
  {NoStop}%
\bibitem [{\citenamefont {Damanet}\ \emph
  {et~al.}(2019{\natexlab{b}})\citenamefont {Damanet}, \citenamefont
  {Mascarenhas}, \citenamefont {Pekker},\ and\ \citenamefont
  {Daley}}]{damanet2019}%
  \BibitemOpen
  \bibfield  {author} {\bibinfo {author} {\bibfnamefont {F.}~\bibnamefont
  {Damanet}}, \bibinfo {author} {\bibfnamefont {E.}~\bibnamefont
  {Mascarenhas}}, \bibinfo {author} {\bibfnamefont {D.}~\bibnamefont {Pekker}},
  \ and\ \bibinfo {author} {\bibfnamefont {A.~J.}\ \bibnamefont {Daley}},\
  }\href@noop {} {\bibfield  {journal} {\bibinfo  {journal} {New Journal of
  Physics}\ }\textbf {\bibinfo {volume} {21}},\ \bibinfo {pages} {115001}
  (\bibinfo {year} {2019}{\natexlab{b}})}\BibitemShut {NoStop}%
\bibitem [{\citenamefont {Sekino}\ \emph {et~al.}(2020)\citenamefont {Sekino},
  \citenamefont {Tajima},\ and\ \citenamefont
  {Uchino}}]{PhysRevResearch.2.023152}%
  \BibitemOpen
  \bibfield  {author} {\bibinfo {author} {\bibfnamefont {Y.}~\bibnamefont
  {Sekino}}, \bibinfo {author} {\bibfnamefont {H.}~\bibnamefont {Tajima}}, \
  and\ \bibinfo {author} {\bibfnamefont {S.}~\bibnamefont {Uchino}},\ }\href
  {\doibase 10.1103/PhysRevResearch.2.023152} {\bibfield  {journal} {\bibinfo
  {journal} {Phys. Rev. Research}\ }\textbf {\bibinfo {volume} {2}},\ \bibinfo
  {pages} {023152} (\bibinfo {year} {2020})}\BibitemShut {NoStop}%
\bibitem [{\citenamefont {Ono}\ \emph {et~al.}(2021)\citenamefont {Ono},
  \citenamefont {Higomoto}, \citenamefont {Saito}, \citenamefont {Uchino},
  \citenamefont {Nishida},\ and\ \citenamefont {Takahashi}}]{ono2021}%
  \BibitemOpen
  \bibfield  {author} {\bibinfo {author} {\bibfnamefont {K.}~\bibnamefont
  {Ono}}, \bibinfo {author} {\bibfnamefont {T.}~\bibnamefont {Higomoto}},
  \bibinfo {author} {\bibfnamefont {Y.}~\bibnamefont {Saito}}, \bibinfo
  {author} {\bibfnamefont {S.}~\bibnamefont {Uchino}}, \bibinfo {author}
  {\bibfnamefont {Y.}~\bibnamefont {Nishida}}, \ and\ \bibinfo {author}
  {\bibfnamefont {Y.}~\bibnamefont {Takahashi}},\ }\href@noop {} {\bibfield
  {journal} {\bibinfo  {journal} {Nature Communications}\ }\textbf {\bibinfo
  {volume} {12}},\ \bibinfo {pages} {6724} (\bibinfo {year}
  {2021})}\BibitemShut {NoStop}%
\bibitem [{\citenamefont {Yamamoto}\ \emph {et~al.}(2021)\citenamefont
  {Yamamoto}, \citenamefont {Nakagawa}, \citenamefont {Tsuji}, \citenamefont
  {Ueda},\ and\ \citenamefont {Kawakami}}]{PhysRevLett.127.055301}%
  \BibitemOpen
  \bibfield  {author} {\bibinfo {author} {\bibfnamefont {K.}~\bibnamefont
  {Yamamoto}}, \bibinfo {author} {\bibfnamefont {M.}~\bibnamefont {Nakagawa}},
  \bibinfo {author} {\bibfnamefont {N.}~\bibnamefont {Tsuji}}, \bibinfo
  {author} {\bibfnamefont {M.}~\bibnamefont {Ueda}}, \ and\ \bibinfo {author}
  {\bibfnamefont {N.}~\bibnamefont {Kawakami}},\ }\href {\doibase
  10.1103/PhysRevLett.127.055301} {\bibfield  {journal} {\bibinfo  {journal}
  {Phys. Rev. Lett.}\ }\textbf {\bibinfo {volume} {127}},\ \bibinfo {pages}
  {055301} (\bibinfo {year} {2021})}\BibitemShut {NoStop}%
\bibitem [{\citenamefont {Krinner}\ \emph {et~al.}(2017)\citenamefont
  {Krinner}, \citenamefont {Esslinger},\ and\ \citenamefont
  {Brantut}}]{krinner2017}%
  \BibitemOpen
  \bibfield  {author} {\bibinfo {author} {\bibfnamefont {S.}~\bibnamefont
  {Krinner}}, \bibinfo {author} {\bibfnamefont {T.}~\bibnamefont {Esslinger}},
  \ and\ \bibinfo {author} {\bibfnamefont {J.-P.}\ \bibnamefont {Brantut}},\
  }\href@noop {} {\bibfield  {journal} {\bibinfo  {journal} {Journal of
  Physics: Condensed Matter}\ }\textbf {\bibinfo {volume} {29}},\ \bibinfo
  {pages} {343003} (\bibinfo {year} {2017})}\BibitemShut {NoStop}%
\end{thebibliography}
%

\end{document}